\documentclass{article}

\usepackage{graphics}
\usepackage{epsfig}

\newcommand{\half}{\frac{1}{2}}

\begin{document}

%%%%%%%%%%%%%%%%%%%%% The title-page %%%%%%%%%%%%%%%%%%%%%%%%%%%%%%%%%%%%%%%%%
\begin{titlepage}
\begin{flushright}
LU TP 01-22 \\
June 2001
\end{flushright}
\begin{center}
\Large
{\bf The Lund  Fragmentation Process for a Multi-gluon  String According
to the Area Law} \\
\normalsize
\vspace{2mm}
Bo Andersson\footnote{bo@thep.lu.se}  \\
Sandipan Mohanty\footnote{sandipan@thep.lu.se}  \\
Fredrik S\"oderberg\footnote{fredrik@thep.lu.se} \\

Department of Theoretical Physics, Lund University, \\
S\"olvegatan 14A, S-223 62 Lund, Sweden
\end{center}
\noindent  {\bf   Abstract}  \\  The  Lund  Area   Law  describes  the
probability for the production of  a set of colourless hadrons from an
initial set  of partons, in  the Lund string fragmentation  model.  It
was  derived from classical  probability concepts  but has  later been
interpreted as the  result of gauge invariance in  terms of the Wilson
gauge loop  integrals.  In this paper  {\it we will  present a general
method to implement  the Area Law for a  multi-gluon string state}. In
this case the  world surface of the massless  relativistic string is a
geometrically  bent   $(1+1)$-dimensional  surface  embedded   in  the
$(1+3)$-dimensional  Minkowski  space.  The  partonic  states  are  in
general  given by  a  perturbative QCD  cascade  and are  consequently
defined   only   down   to   a   cutoff   in   the   energy   momentum
fluctuations. {\it  We will  show that our  method defines  the states
down  to the  hadronic mass  scale inside  an  analytically calculable
scenario}.

{\it We  will then show  that there is  a differential version  of our
process  which is closely  related to  the generalised  rapidity range
$\lambda$,  which  has  been  used   as  a  measure  on  the  partonic
states}. We  identify $\lambda$ as  the area spanned between  {\it the
directrix  curve}  (the curve  given  by  the  parton energy  momentum
vectors laid out in colour order, which determines the string surface)
and  the average  curve (to  be called  the ${\cal  P}$-curve)  of the
stochastic   $X$-curves    (curves   obtained   when    the   hadronic
energy-momentum vectors are laid out  in rank order).  Finally {\it we
show that from the  $X$-curve corresponding to a particular stochastic
fragmentation  situation it  is  possible to  reproduce the  directrix
curve} (up to  one starting vector and a set of  sign choices, one for
each hadron).  This relationship provides an analytical formulation of
the notion of Parton-Hadron Duality. The whole effort is made in order
to get  a new handle to  treat the transition region  between where we
expect perturbative QCD to work and where the hadronic features become
noticeable.

\end{titlepage}

%%%%%%%%%%%%%%%%%%%%%%%%%%%%%%%%%%%%%%%%%%%%%%%%%%%%%%%%%%%%%%%%%%%%%%%%%%%%%%%%%

\section{Introduction}
The  Lund String  Fragmentation Model  was developed  many  years ago,
\cite{BAGGBS} ,\cite{BA},  and as implemented in  the well-known Monte
Carlo  simulation   program  JETSET,  \cite{TS},  it   has  been  very
successful   in  reproducing  experimental   data  from   high  energy
multi-particle processes.

\begin{figure}
\begin{center}
\includegraphics{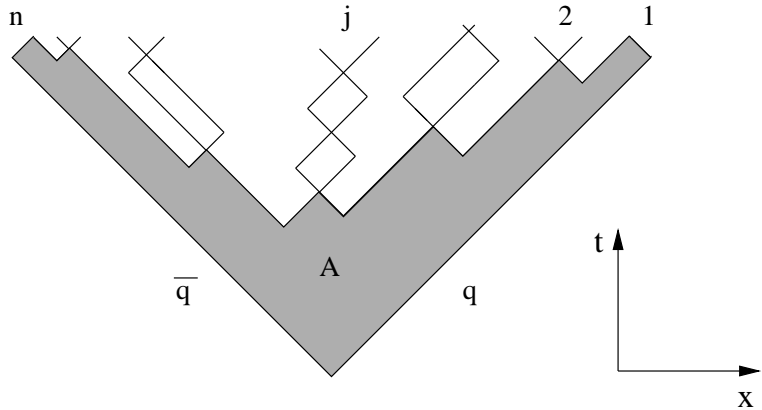}
\end{center}
\caption{A high-energy string breakup.} \label{area}
\end{figure}

The model is based upon a few general assumptions: (i) the final state
particles stem from  the breakup of a string-like  force field spanned
between the coloured constituents, (ii) there is causality and Lorentz
invariance and (iii) the production  of the particles can be described
in terms of a stochastic  process which obeys a saturation assumption.
We have,  in a recent  paper \cite{BAFS}, re-derived the  major result
for the $(1+1)$-dimensional model, which is applicable for events with
a  quark   ($q$,  a  colour-$3$)   and  an  antiquark   ($\bar{q}$,  a
colour-$\bar{3}$) at the endpoints of  the string but with no interior
gluonic  (g,   colour-$8$)  excitations.   The  result  is   that  the
(non-normalised)  probability for  the production  of  an $n$-particle
final  state  of hadrons  with  energy  momenta  $\{p_j\}$ and  masses
$\{m_j\}$ is given by the Lund Area Law:

\begin{eqnarray}
\label{arealaw}
dP_n(\{p_j\};P_{tot})=\prod_{j=1}^n   N_j  d^2p_j  \delta(p_j^2-m_j^2)
\delta( \sum_{j=1}^n p_j - P_{tot}) \exp(-b A)
\end{eqnarray}

where $ A$  is the area spanned by the  string ``before'' the breakup,
cf.   Fig.\ref{area}, $P_{tot}$  is the  total energy-momentum  of the
state and $\{N_j\}$  and $b$ are parameters related  to the density of
hadronic  states  and the  breakup  properties  of  the string  field,
respectively.

The result  in Eq. (\ref{arealaw})  is evidently similar to  a quantum
mechanical transition  probability. It is the final  state phase space
multiplied  by a  squared matrix  element,  which in  this case  would
correspond to  the negative area  exponential. In \cite{BAFS}  we have
shown  that it  is  possible  to ``diagonalise''  the  model, i.e.  to
express the result in Eq.   (\ref{arealaw}) as a product of (diagonal)
transition  operators   (in  quantum  mechanical   language  it  would
correspond to  density operators). It turns  out that in  this way the
dynamics can be described in terms of $(1+1)$-dimensional (space-like)
harmonic oscillators.

Shortly after the original  derivation of the Area Law, \cite{BAGGBS},
Sj\"ostrand,  \cite{TSII},  provided an  implementation  of the  model
applicable also  for multi-gluon states, i.e. when  the string surface
is  no  longer  flat  but  geometrically  bent  due  to  the  internal
excitations. Sj\"ostrand's  method is to project the  positions of the
breakup  points  (the vertices)  from  the (flat)  $(1+1)$-dimensional
model as  given by  Eq. (\ref{arealaw}) onto  the surface of  the bent
string.   The projection  is  done so  that  the proper  times of  the
vertices and the squared masses of the particles produced between them
are the same.  Unfortunately this  method does not fulfil the Area Law
on the  bent surface because it  is a geometrical fact  that the areas
``below the vertices'' are not  invariant under such a projection from
a flat to a bent surface.

Although the Area  Law is not fulfilled on an event  to event basis by
the method  in \cite{TSII}  we will  show that it  is fulfilled  in an
average sense, i.e.  the  predicted inclusive distributions are little
affected by  the differences.  It is well-known  that the experimental
results for these  distributions are well described by  JETSET even up
to the largest available energies of today.

The intention of this note is to implement another method for particle
production in multi-gluon  states which fulfils the Area  Law at every
single  step  in the  production  process. We  will  find  that it  is
necessary to tackle a set of  problems in the definition of the states
which we  apply the  process to.  We note that  the states  defined by
perturbation theory are resolved only  to the scale of some virtuality
cutoff. We will find that our  method provides a set of excitations on
the hadronic mass scale, in the string field.  We will investigate the
properties of these ``soft hadronisation gluons'' in future work.

The  states  of the  massless  relativistic  string  fulfil a  minimum
principle, i.e. the surface spanned  in spacetime by the string during
its motion is a minimal surface.  This means on the one hand, that the
states  should be  stable against  small-scale variations  and  on the
other  hand that  the  surface  is fully  determined  by the  boundary
curve. In this case the boundary curve corresponds to the orbit of one
of   the  endpoints,  conventionally   chosen  as   the  $q$-endpoint.
Therefore, the process we are going  to define is a process along this
curve, to  be called  the {\it directrix  curve}, which  is completely
defined by the  perturbative cascade. In this paper  we will treat the
partons as massless, although both the process and the directrix curve
can be defined for a general case with massive quarks.

One property which can be derived from Eq. (\ref{arealaw}) is that the
average decay region is a typical hyperbola. On the average, the final
state hadrons in our process will  be produced in the same way, albeit
this time along  a set of connected hyperbolae.  In \cite{BAGGBSII} we
have defined such an average curve and we will, in this paper, call it
the  ${\cal  X}$-curve.  Just  as  a simple  hyperbola  has  a  length
proportional to  the hyperbolic angle that it  spans (this corresponds
to  the available rapidity  range along  the mean  decay region,  in a
two-jet  system  of  hadrons)   the  ${\cal  X}$-curve  has  a  length
corresponding  to a  generalised  rapidity range,  usually called  the
$\lambda$ measure, \cite{BAGGBSII}, \cite{BA}. The ${\cal X}$-curve is
defined in terms of differential  equations and we will show the close
relationship between the ${\cal X}$-curve and our process in the limit
of a vanishing hadron mass.

There are several  reasons to undertake this investigation.  One is to
compare the precise implementation of  the Area Law to the approximate
process in  \cite{TSII}. We  will do  this both in  this paper  and in
future publications.

Another  reason  is  to get  a  handle  on  the general  structure  of
fragmentation, in particular to be  able to treat also the multi-gluon
fragmentation   states  by  the   analytical  methods   introduced  in
\cite{BAFS}. This is of particular interest for the transition region,
i.e. the region in between where we expect perturbation theory to work
and where we know that the non-perturbative fragmentation sets in.

A final  reason is to investigate  the stability of the  states in QCD
under fragmentation, i.e. given  a multi-gluon state defined according
to the rules of perturbation theory (with cut-offs as mentioned above)
to find out  to what extent it can be modified  so that the observable
results  after   fragmentation  are   still  in  agreement   with  the
experiments.  In the  Lund interpretation  of fragmentation  where the
particles  stem from  the  energy of  the  force field  it is  tacitly
assumed that modifications  of the perturbative state below  and up to
the scale of the hadronic masses  should have no effects. We will find
that it is necessary to  take into account the coherence properties of
the radiation in any modification.

Although the  methods presented  in this paper  are applicable  to any
multi\-gluon state, for definiteness we will concentrate on the states
obtained  in  $e^+e^-$-annihi\-lation  processes  where we  expect  an
original colour  singlet $(q \bar{q})$-state  to form and start  to go
apart, producing a set  of bremsstrahlung gluons inside an essentially
point-like region. We will also be satisfied to treat a single kind of
hadron with  mass $m$.  Finally, in this  paper we will  not introduce
gaussian  transverse  momentum  fluctuations  (which we  expect  in  a
tunneling scenario  \cite{BA}) in  the fragmentation process.  We will
investigate   the  influence   of  such   fluctuations  in   a  future
publication.

In  Section \ref{lundmodel}, we  provide a  set of  necessary formulae
from the  $(1+1)$-dimensional Lund  Model.  In Section  \ref{3dim}, we
consider  the motion  of strings  containing internal  excitations. We
also provide a description of the coherence properties of QCD and some
necessary  formulae to  understand  the ${\cal  X}$-curve. In  Section
\ref{singlepoint}, we  define the properties of the  string breakup in
general. We find that the  most general implementation of the Area Law
leads  to  a  highly  intractable  process and  therefore  in  Section
\ref{method2} we define another  approach which mends all the problems
with the  earlier one.  In  particular, in Section  \ref{diffcurve} we
show  a close  relationship  between a  differential  version of  this
process  and  the  ${\cal  X}$-curve.  In  Section  \ref{results},  we
present a set of results  from our method, followed by some concluding
remarks on future work in section 7.

\pagebreak
\section{Some results from the $(1+1)$-dimensional model}
\label{lundmodel}

\subsection{The Area Law}
\label{AreaLaw}

The  Lund  Model contains  a  non-trivial  interpretation  of the  QCD
force-field  in terms  of the  massless relativistic  string  with the
quarks ($q$) and the anti-quarks  ($\bar{q}$) at the endpoints and the
gluons  ($g$) as  internal  excitations  on the  string  field. It  is
assumed that  the force field can  break up into smaller  parts in the
fragmentation process  by the  production of new  ($q \bar{q}$)-states
(i.e. new endpoints). A $q$  from one such breakup point (``vertex''),
together  with a  $\bar{q}$  from  an adjacent  vertex  and the  field
between them, can form a hadron on mass shell.

For the  simple case when there  are no gluons, the  string field only
corresponds to  a constant force  field (with a  phenomenological size
$\kappa   \simeq  1   GeV/fm$)   spanned  between   the  original   $q
\bar{q}$-pair.   In   a   semi-classical  picture,   conservation   of
energy-momentum allows  the creation  of a new  massless pair  at some
point  along the field.  The pair  will then  go apart  along opposite
light-cones, using up the energy in  the field in between (in this way
the confined fields will always end on the charges). In order that the
hadron produced  between two adjacent vertices should  have a positive
squared  mass, it  is  necessary that  the  vertices are  placed in  a
space-like   manner  with  respect   to  each   other.   Consequently,
time-ordering  will be  a  frame-dependent statement  (in any  Lorentz
frame the slowest particles will  be the first to be produced, thereby
fulfilling  the requirements  in a  Landau-Pomeranchuk  formation time
scenario).  It is possible  to order  the production  process instead,
along the light-cones  and introduce the notion of  {\it rank} so that
the first  rank hadron along the original  $q$-light-cone will contain
that $q$  together with  a $\bar{q}$ from  the first vertex  along the
light-cone, the second  rank hadron a $q$ from the  first vertex and a
$\bar{q}$ from the  next etc., cf. Fig.\ref{area}. Rank  ordering is a
frame independent procedure.  It is  of course possible to introduce a
rank-ordering also from the end containing the original $\bar{q}$.

One  obtains  ,  \cite{BAGGBS},  \cite{BA},  \cite{BAFS},  the  unique
process described  by Eq. (\ref{arealaw}) from  these observations and
an assumption that the  breakup process obeys a saturation assumption,
i.e. that  after many  steps when  we are far  from the  endpoints the
proper  times of  the vertices  will  be distributed  according to  an
energy-independent distribution.

A  particular  feature is  that  if  a  particle with  energy-momentum
$p=(p_+,p_-)$  and  with  squared  mass $m^2=p^2=p_+p_-$  is  produced
between     the     two     vertices    with     $x=(x_+,x_-)$     and
$x^{\prime}=(x_+^{\prime},x_-^{\prime})$       then       we      have
(cf. Fig.\ref{1particlea}):

\begin{figure}
\begin{center}
\includegraphics{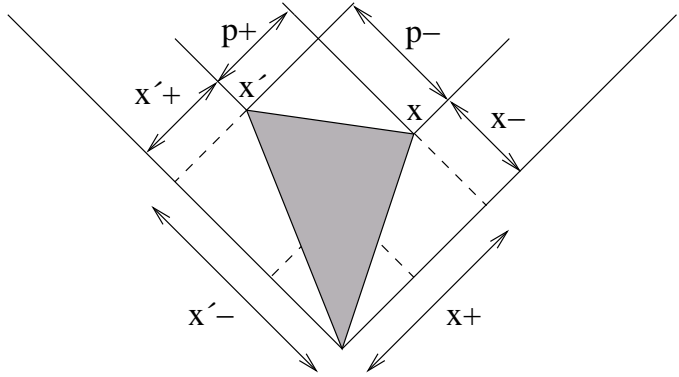}
\end{center}
\caption{Two  adjacent  vertices   of  the  string  breakup  process.}
\label{1particle} \label{1particlea}
\end{figure}

\begin{eqnarray}
\label{energymomentum}
p_+= \kappa (x_+-x_+^{\prime})\equiv q_+-q^{\prime}_+ \nonumber \\
p_-= \kappa (x_-^{\prime}-x_-) \equiv q_--q^{\prime}_-
\end{eqnarray}
Thus we find that on a  flat string surface the difference between the
vertex points will fulfil :
\begin{eqnarray}
\label{vertexdiff}
(x-x^{\prime})^2=-m^2/\kappa^2
\end{eqnarray}
Eq. (\ref{energymomentum})  implies that the  $(1+1)$-dimensional Lund
Fragmentation   Model  may   also   be  described   by   means  of   a
multi-peripheral  chain diagram  as  in Fig.\ref{multipera}  or in  an
energy-momentum space picture as in Fig.\ref{multiperb}.

\begin{figure}
\begin{center}
\includegraphics{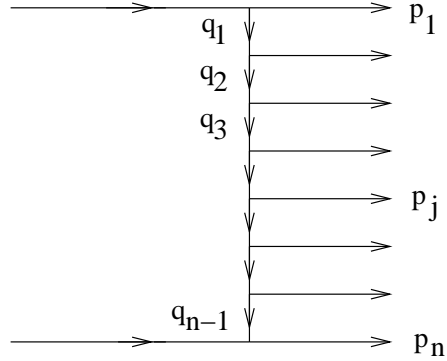}
\end{center}
\caption{A  ladder   diagram  describing  multi-particle  production.}
\label{multipera}
\end{figure}
\begin{figure}
\begin{center}
\includegraphics{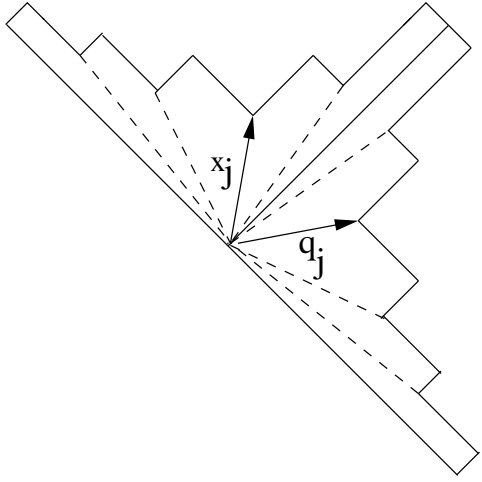}
\end{center}
\caption{The energy-momentum space picture ($q_{j}$) and the spacetime
picture ($x_{j}$) respectively.} \label{multiperb}
\end{figure}

This is used in \cite{BAFS} in order to subdivide the Area Law process
into  steps in  between the  vertices. The  energy-momentum conserving
$\delta$-distribution  in  Eq. (\ref{arealaw})  can  be ``solved''  by
introducing  the momentum  transfers $\{q_j\}$  instead of  the hadron
momenta  $\{p_j\}$.   Then the  mass-shell  condition  means that  the
hyperbolic angle between  the vertices and the size  of the area slit,
exhibited in Fig.\ref{1particlea} are  fixed by the squared sizes $q^2
=-\Gamma$,            $(q^{\prime})^2=-\Gamma^{\prime}$            and
$(q-q^{\prime})^2=m^2$. The result is  that Eq. (\ref{arealaw}) can be
rewritten as a product of steps between the $\{\Gamma_j\}$:

\begin{eqnarray}
\label{steps}
dP_n(\{p_j\},P_{tot})  =  \prod  K(\Gamma_j,\Gamma_{j-1},m^2)d\Gamma_j
\nonumber \\  K(\Gamma,\Gamma^{\prime},m^2)= N\frac{ \exp(-\frac{b}{2}
\sqrt{\lambda(\Gamma,
\Gamma^{\prime},-m^2)})}{\sqrt{\lambda(\Gamma,\Gamma^{\prime},-m^2)}}\nonumber\\
\lambda(a,b,c)=a^2+b^2+c^2-2ab-2ac-2bc
\end{eqnarray}
It  is  a remarkable  fact  that the  transfer  operators  $K$ can  be
diagonalised in terms of the eigenfunctions of the harmonic oscillator
(those which  are boost-invariant in  a $(1+1)$-dimensional space-like
Minkowski space. in a  two-dimensional Euclidean space they correspond
to a  vanishing angular  momentum) $g_n(\Gamma)$ with  the eigenvalues
solely  determined by  the squared  mass  of the  hadrons produced  in
between:
\begin{eqnarray}
\label{diagonal}
K(\Gamma,\Gamma^{\prime},m^2)=\sum_{n=0}^{\infty}g_n(\Gamma)\lambda_n(m^2)
g_n(\Gamma^{\prime})
\end{eqnarray}
The eigenvalues $\lambda_n$ are analytic continuations of the harmonic
oscillator  eigenfunctions  to   time-like  values  of  the  argument,
\cite{BAFS}.   Useful  representations  of  $K$  and  the  eigenvalues
$\lambda_n$ are :
\begin{eqnarray}
K(\Gamma,\Gamma^{\prime},m^2)=     \int_0^1     \frac{dz}{z}    \exp(-
\frac{b}{2}(z\Gamma+\frac{m^2}{z}))\delta                             (
\Gamma^{\prime}-(1-z)(\Gamma+\frac{m^2}{z}))                \nonumber\\
\lambda_n(m^2)=N   \exp(\frac{bm^2}{2})\int_0^1  \frac{dz}{z}  (1-z)^n
\exp(-\frac{bm^2}{z})
\end{eqnarray}
We have  introduced the positive  light-cone fraction of  the produced
hadron  $z$  defined  by  $(x_+-x^{\prime}_+)=zx_+$.  It  is  straight
forward  algebra   to  prove  that  the  shaded   area,  exhibited  in
Fig.\ref{1particle}  is  given  by the  exponent  $\frac{1}{2}(z\Gamma
+\frac{m^2}{z})$  in the representation  of the  kernel $K$.   We also
note that the area obtained by summing the areas of the regions marked
$I$, $II$  and $III$  in Fig.\ref{1particleb} is  equal to  twice this
area and that  $m^2/z$ equals the sum of the areas  of regions $I$ and
$II$. There is  a simple relationship between the  two adjacent values
of $\Gamma$ in the representation of $K$:

\begin{figure}
\begin{center}
\includegraphics{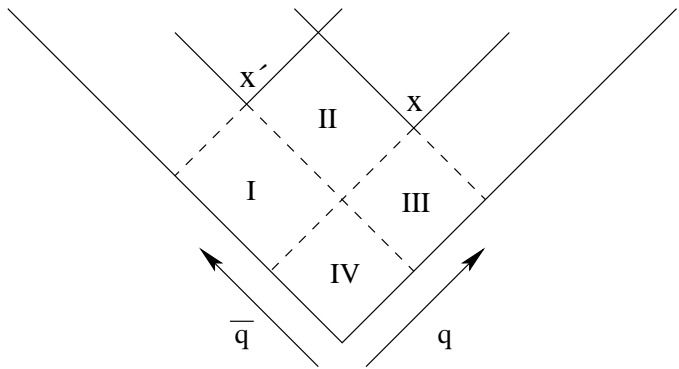}
\end{center}
\caption{The figure shows the two  vertices and the regions I, II, III
and IV described in the text.} \label{1particleb}
\end{figure}

\begin{eqnarray}
\label{Gammarel}
\Gamma^{\prime}=(1-z)(\Gamma+\frac{m^2}{z})
\end{eqnarray}
 Finally we note the identity (in easily understood notation) :
\begin{eqnarray}
\label{nonlinear}
(I)(III)=(II)(IV)
\end{eqnarray}
Eq. (\ref{nonlinear}),  just as Eq. (\ref{vertexdiff}),  is only valid
for a  flat string  surface and a  particular consequence is  that the
region $IV$  is non-vanishing  for the $(1+1)$-dimensional  model.  As
the area of  the region $IV$ is proportional  to $(1-z)$, the variable
$z$  must always  be smaller  than  unity, i.e.  there is  a built  in
requirement that  a typical step of  the process can never  use up all
the available light-cone energy-momentum.

Such a requirement  also comes out of the  following argument. Suppose
that we would integrate $dP_n$ in Eq (\ref{arealaw}) over all possible
energy momenta  and then sum  over all multiplicities. Due  to Lorentz
invariance, we  will obtain  a function $R(s)$  which can  only depend
upon the  total squared energy-momentum $s=P^2_{tot}$. If  we pick out
the dependence  on the first particle  and sum and  integrate over all
the rest we obtain an integral equation for the function $R$ :
\begin{eqnarray}
\label{Req}
R(s)=   (B.T)   +   \int_0^1  N   \frac{dz}{z}   \exp(-b\frac{m^2}{z})
R(s^{\prime})\nonumber\\ s^{\prime}=(1-z) (s-\frac{m^2}{z})
\end{eqnarray}
where $(B.T)$ stands for  ``boundary condition term'' and the variable
$s^{\prime}$  is  equal to  the  squared  mass  of all  the  remaining
particles if  the first hadron  takes the light-cone fraction  $z$ (we
note  the  similarity  to  the  Eq.  (\ref{Gammarel})).  The  integral
equation (\ref{Req}) has an  asymptotic solution $R \propto s^a$ (with
the  parameter   $a$  being  a   function  of  $N$  and   $bm^2$  ,cf.
\cite{BAFS}) with the requirement :
\begin{eqnarray}
\label{aint}
\int_0^1 N \frac{dz}{z} (1-z)^a \exp(-b\frac{m^2}{z})=1
\end{eqnarray}
Consequently  while the  exclusive  formula for  the  production of  a
particular hadron  with the  light-cone fraction $z$  is given  by the
Area   Law,  the   inclusive  probability   to  produce   this  hadron
(irrespective of what comes after  it in the process) must be weighted
with  $R(s^{\prime})/R(s)\simeq  (1-z)^a$.  Therefore, the  well-known
Lund   fragmentation   formula   is   given  by   the   integrand   in
Eq. (\ref{aint}) and there is  a power suppression for large values of
the fragmentation variable.

The  formulae presented  above  correspond to  an  ordering along  the
positive light-cone, i.e. the variable  $z$ is defined as the positive
light-cone  momentum  fraction of  the  particle.  It  is possible  to
redefine  everything  in  terms  of  an ordering  along  the  negative
light-cone,  i.e. to introduce  the corresponding  negative light-cone
component    $\zeta$    by    writing    $(x^{\prime}_--x_-)    =\zeta
x^{\prime}_-$. It is straightforward to prove that
\begin{eqnarray}
\label{zeta}
\zeta=\frac{m^2}{m^2+z\Gamma}  &  \mbox{and} &  z=\frac{m^2}{m^2+\zeta
\Gamma^{\prime}}
\end{eqnarray}
and from this we find that  the integrand in the representation of the
kernel   $K$   can   be   reformulated   from   $(z,\Gamma)\rightarrow
(\zeta,\Gamma^{\prime})$   to  exhibit   the   symmetry  between   the
descriptions along the positive and the negative light-cone directions
\begin{eqnarray}
\frac{dz}{z}        \delta(\Gamma^{\prime}-(1-z)(\Gamma+\frac{m^2}{z}))
\exp(-\frac{b}{2}(z\Gamma+   \frac{m^2}{z}))  \rightarrow  \nonumber\\
\frac{d\zeta}{\zeta}\delta(\Gamma-(1-\zeta)(\Gamma^{\prime}+
\frac{m^2}{\zeta}))
\exp(-\frac{b}{2}(\zeta\Gamma^{\prime}+\frac{m^2}{\zeta}))
\end{eqnarray}

\section{The description of a multi-gluon string state}
\label{3dim}

The dynamics  of the  massless relativistic string  is based  upon the
principle  that {\it  the surface  spanned  by the  string during  its
motion  is a minimal  surface}. This  means that  {\it the  surface is
completely determined by its boundary}.   In the Lund Model the string
is used as a model for the  confined colour force field in QCD and the
above  property then has  the further  important implication  that the
dynamics will be infrared  stable, i.e.  {\it all predictable features
from  the decay  of the  force field  should be  stable  against minor
deformations of the boundary}.

For an  open string a single  wave moves across  the spacetime surface
and bounces at the endpoints. {\it  The wave motion is determined by a
(four-)vector-valued shape function, which we will call the directrix,
${\cal A}$}. Thus a point on the string, parametrized by the amount of
energy  $\sigma$,  between  the   point  and  (for  definiteness)  the
$q$-endpoint is, at the time $t$, at the position
\begin{eqnarray}
\label{xpoint}
x(\sigma,t)   =   \frac{1}{2}({\cal  A}(t+\frac{\sigma}{\kappa})+{\cal
A}(t-\frac{\sigma}{\kappa}))
\end{eqnarray}
We will from now on put the string constant $\kappa$ equal to unity in
order to simplify the formulae.

While   the  tension   $\vec{T}=\partial\vec{x}/\partial   \sigma$  is
directed       along        the       string,       the       velocity
$\vec{v}=\partial\vec{x}/\partial t$ is  directed transversely so that
$\vec{T}  \cdot \vec{v}=0$.  The definition  of $\sigma$  also implies
that $\vec{T}^2+\vec{v}^2=1$ (all the three-vector relations are valid
in  the local  rest-frame).  Together this  means  that the  directrix
function has an everywhere light-like tangent
\begin{eqnarray}
\label{Adot}
\left(\frac{d   \vec{{\cal  {A}}}}{d   \xi}\right)^2=1=\left(  \frac{d
{\cal{A}}_{0}}{d \xi}\right)^2
\end{eqnarray}
The   tension   must  vanish   at   the   endpoints  ($\sigma=0$   and
$\sigma=E_{tot}$)  and  this implies  that  the  directrix  must be  a
periodic function with the property
\begin{eqnarray}
\label{Aperiod}
{\cal A}(\xi + 2 E_{tot})= {\cal A}(\xi) + 2 P_{tot}
\end{eqnarray}
where $P_{tot}$  ($E_{tot}$) is the total  energy-momentum (energy) of
the state. While according to Eq. (\ref{xpoint}), the directrix ${\cal
A}(t)$ describes the motion  of the $q$-end, from Eq.  (\ref{Aperiod})
it is evident that
\begin{eqnarray}
\label{barq}
{\cal A}_{\bar{q}}(t)={\cal A}(t+E_{tot})- P_{tot}
\end{eqnarray}
will describe the motion of  the $\bar{q}$-end. Finally, if the string
starts out  from a  point (at the  time $t=0$)  then we must  have the
symmetry :
\begin{eqnarray}
\label{Asym}
{\cal A}(\xi)=-{\cal A}(-\xi)
\end{eqnarray}
Using the Lund interpretation of the gluons as internal excitations on
the  string it  is easy  to  construct the  first half  period of  the
directrix curve:  it starts with  the quark energy-momentum  $k_1$ and
then the gluon  energy-momenta $\{k_j\}$ are laid out  in colour order
and it  ends with  the $\bar{q}$ energy-momentum  $k_n$, for  a string
with $(n-2)$ gluons.  In this  way the $q$-endpoint will be acted upon
by  the  colour-ordered excitations  as  they  arrive  in turn.   From
Eqs. (\ref{Aperiod})  and (\ref{Asym}) it  follows that we  obtain the
directrix of the  second half period by reversing  the order, starting
with   the  $\bar{q}$   energy-momentum  and   ending  with   the  $q$
energy-momentum  (besides   the  translation  this  is   the  way  the
$\bar{q}$-endpoint will move according to Eq. (\ref{barq})).

The  energy-momentum content  in the  string  at a  certain time  $t$,
between the point $\sigma$ and the $q$-end is given by :
\begin{eqnarray}
\label{momentum}
\int_0^{\sigma}  d\sigma^{\prime}   \frac{\partial  x}{\partial  t}  =
\frac{1}{2} ({\cal A}(t+\sigma)-{\cal A}(t-\sigma))
\end{eqnarray}

\subsection{The coherence conditions in QCD}
\label{cohQCD}

The properties of the directrices which are described above are common
to all states  of the massless relativistic string.  On the other hand
the use  of the  string as  a model for  the force  fields of  QCD may
single out a particular class  of all possible states. We will briefly
discuss  {\it  the  conditions   which  correspond  to  the  coherence
conditions of bremsstrahlung radiation in a gauge field theory}.

Multi-gluon radiation is in general described by means of perturbative
cascade models. In order to  consider the properties of the states, we
will make  use of the ideas behind  the Lund Dipole Model.  To see the
emergence of a  multi-gluon state we start out  with the following two
basic results from  QCD bremsstrahlung, (cf. \cite{BA}, \cite{YDVKST},
\cite{GG}).

\begin{itemize}

\item[B1]  The  original   $(q  \bar{q})$-state  emits  bremsstrahlung
radiation according to  the ordinary dipole formula, i.e.  there is an
inclusive density  of gluon quanta which  is flat in  rapidity $y$ and
the   logarithm   of   the   squared  transverse   momentum   $\kappa=
\log(k_{\perp}^2)$   with   a   density   given   by   the   coupling,
$\bar{\alpha}\equiv C\alpha_s/2\pi$:
\begin{eqnarray}
\label{dipole1}
 dn= \bar{\alpha} d\kappa dy
\end{eqnarray}

\item[B2] After the first (``hardest'') gluon ($g_1$) is emitted there
are two  dipoles available, one  between the $(qg_1)$ and  one between
the  $(g_1\bar{q})$   with  squared  masses  $s_{12}   \equiv  (k_q  +
k_{g_1})^2$  and  $s_{23}\equiv  (k_{g_1}+k_{\bar{q}})^2$. The  second
gluon may be emitted independently inside the angular (rapidity) range
of the first or the second  dipole. We note that the two dipoles (each
spanned by  two light-like  parton energy momenta)  move apart  in the
total cms.

\end{itemize}
After the emission  of the second gluon, according  to the Lund Dipole
Model  \cite{GG},  there are  three  independent  dipoles for  further
emission, and so on.  The  masses of the dipoles will quickly decrease
and  therefore  also  the  transverse  momentum size  of  the  emitted
gluons. The cascades are stopped by some ``virtuality'' cutoff, either
in the dipole mass or in  the transverse momentum. The Dipole Model is
implemented in the Monte Carlo simulation program ARIADNE \cite{LL}.

\begin{figure}
\begin{center}
\includegraphics{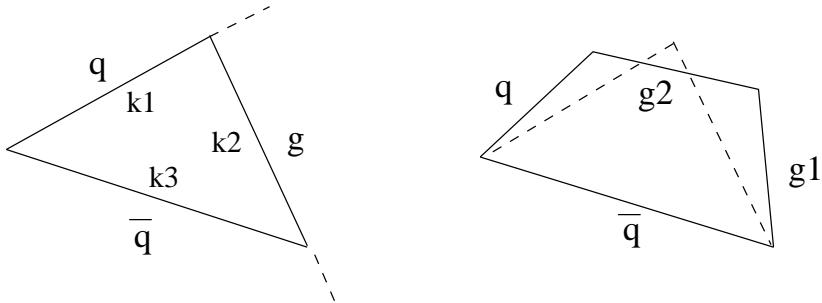}
\end{center}
\caption{The emission of the first and the second gluon respectively.}
\label{3part}
\end{figure}

In  terms of the  directrix, the  original state  is described  by the
light-like energy momenta of the original $(q\bar{q})$-pair. After the
emission of the first gluon the  state, in space (in the cms), will be
described by  a (connected)  triangle, cf. Fig.\ref{3part},  where the
vectors $\vec{k}_q\equiv  \vec{k}_1$ and the  vectors $\vec{k}_g\equiv
\vec{k}_2$  constitute one  of  the dipoles  and  the $\vec{k}_2$  and
$\vec{k}_{\bar{q}}\equiv  \vec{k}_3$ the  second dipole,  as discussed
above.  (We note that for  light-like vectors the space part length is
equal to  the time component) Emission  of the second  gluon will then
occur  in between  the  two  vectors describing  the  dipole and  this
evidently means that  the second gluon vector will  ``cut off'' one of
the  triangular  corners,  making  the  directrix  into  a  quadrangle
etc. (Note that  already with the emission of  two gluons, the vectors
shown in Fig.\ref{3part} no longer need to be in a plane)

In general,  there will  be an  emission of a  set of  ``hard'' gluons
which will determine the general shape of the directrix. The remaining
emissions will then  make the directrix smoother and  smoother as each
new  emission will  correspond to  a gluon  vector which  cuts  off an
earlier corner. This means  that the angle between the energy-momentum
vectors of  colour-connected partons  becomes smaller (along  the main
directions, determined  by the hard emissions) the  longer the cascade
continues.  This is  the  way  that the  coherence  properties of  the
bremsstrahlung radiation work, and it  is sometimes referred to as the
``strong angular condition''. Later on, we will find that this angular
condition will play an important  role when we consider the deviations
in the partonic states which are allowed by the fragmentation process.

\section{The General Breakup of a String Field}
\label{singlepoint}

We will now  consider the partitioning of a general  string state at a
point  $x(\sigma,t)$ and after  that define  the most  general process
possible for the Area Law.

\begin{figure}[!t]
\begin{center}
\includegraphics{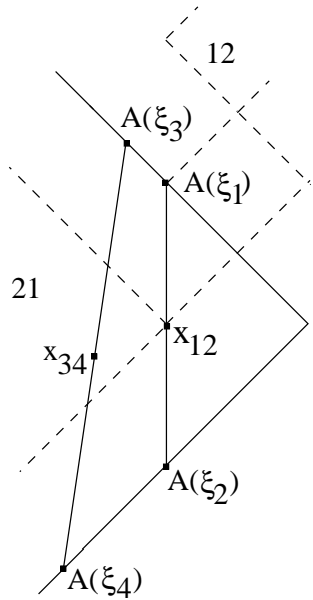}
\end{center}
\caption{ The system  (12) is shown as a string  piece which is moving
away,  while the  system (21)  is translated  to $x_{12}$.  The second
break-up point  $x_{34}$ is the middle point  between $A(\xi_{3})$ and
$A(\xi_{4})$. In  this way, the particle produced  in between $x_{12}$
and  $x_{34}$  can be  taken  either as  the  second  particle in  the
break-up of  the original system or  the first in the  system (21).  }
\label{divide}
\end{figure}

According to Eq.  (\ref{xpoint}) the breakup will occur  at the middle
point $\half({\cal  A}(\xi_1)+{\cal A}(\xi_2))$ between  two positions
on    the     directrix,    determined    by     $\xi_1>\xi_2$    with
$t=\half(\xi_1+\xi_2)$         and        $\sigma=\half(\xi_1-\xi_2)$,
cf. Fig.\ref{divide}.  There will be two parts of the string left over
and we will now describe their motion after the breakup.

The  first part (to  be denoted  ``$(12)$'') can  be described  by the
``new''directrix   ${\cal   A}_{12}(\xi)\equiv   {\cal  A}(\xi)$   for
$\xi_2\leq \xi\leq\xi_1$. It will contain the energy-momentum $P_{12}=
({\cal A}(\xi_1)-{\cal A}(\xi_2))/2$, cf. Fig.\ref{divide}, so that it
can be  continued as in Eq. (\ref{Aperiod})  with $P_{tot} \rightarrow
P_{12}$. Starting it out from the position ${\cal A}_{12}(\xi_2)\equiv
{\cal  A}(\xi_2)$,   it  is  the  ``new''  orbit   of  the  (original)
$q$-particle in the string  $(12)$.  To obtain the corresponding orbit
for the ``new''  $\bar{q}$-end we use the Eq.   (\ref{barq}) to obtain
${\cal             A}_{12             \bar{q}}\equiv             {\cal
A}_{12}(t+(\xi_1-\xi_2)/2)-P_{12}$.  If we  start  at the  ``breakup''
time $t=\half(\xi_1+\xi_2)$,  it will evidently  behave just as  if we
had adjoined the directrix  ${\cal A}_{12}$ to the point $x(\sigma,t)$
(note that ${\cal A}_{12}(\xi_1)= 2P_{12}+ {\cal A}_{12}(\xi_2)$).

The  second  part  (denoted   ''$(21)$'')  can  be  described  as  the
remainder, i.e. ${\cal  A}_{21}={\cal A}(\xi)$ with $\xi_1\leq \xi\leq
2E_{tot}+\xi_2$  (noting  that  the  original directrix  is  continued
according  to Eq.  (\ref{Aperiod}))  . The  directrix ${\cal  A}_{21}$
contains   the  energy   $P_{21}=P_{tot}-P_{12}$   and  is   continued
accordingly. To  find the  orbit of the  produced $q$-end in  the part
$(21)$  we   adjoin  the  directrix  ${\cal  A}_{21}$   to  the  point
$x(\sigma,t)$  starting it at  $\xi=\xi_1$.  The  (original) $\bar{q}$
will again move according to Eq. (\ref{barq}).

The most noticeable property is that for a multi-gluon force field the
directrices   of  the   two   new  string   parts   will  not   fulfil
Eq. (\ref{Asym}). This  implies that the endpoints of  the new strings
never meet but instead turn around  each other so that each of the two
states  will  contain  angular  momentum.   {\it It  is  only  in  the
$(1+1)$-dimensional case that  the two new string parts  will have the
same properties  as the original  string} (though they will  be scaled
down  in size  and  have  different rest-frames  due  to the  momentum
transfer at the breakup).  The conclusion is that {\it the final state
in  a multi-gluon  string fragmentation  in general  depends  upon the
ordering of the different breakups}.

\subsection{A general process based upon the Area Law}
\label{genprocess}

We may nevertheless  devise a string breaking process  on a ``frozen''
string surface, i.e.  under the assumption that the  string surface is
once and  for all  a given  device (this is  also the  way Sj\"ostrand
treated the  problem exhibited above).  Assume that  the system $(12)$
is a hadron on the  mass-shell, produced as the ``first'' hadron along
the original $q$-direction (i.e.  the point ${\cal A}(\xi_2)$ is along
the $q$ energy-momentum vector and ${\cal A}(\xi_1)$ on the other side
of  the  corner  in  the  directrix  between the  $q$  and  the  first
gluon).  Then   we  may  choose  two  new   points  $\xi_3>\xi_1$  and
$\xi_4<\xi_2$   to  obtain   the  second   rank  hadron   $(34)$  with
energy-momentum $p_{34}=\half  ({\cal A}(\xi_3)- {\cal A}(\xi_1)+{\cal
A}(\xi_2)-   {\cal   A}(\xi_4))$  with   the   production  vertex   at
$x_{34}=\half({\cal  A}(\xi_3)+{\cal  A}(\xi_4))$.  We  may  evidently
continue this  process across the string  surface and we  note that at
every  step it  is  only  necessary to  choose  two numbers:  $(\delta
\xi)_+=\xi_3-\xi_1$  and $(\delta  \xi)_-=\xi_2-\xi_4$.  They must  be
chosen so that the corresponding  $p_{34}$ is the energy-momentum of a
particle  on  the  mass-shell.   It  is necessary  to  have  a  second
condition, however, and we may then use the Area Law.

For  the $(1+1)$-dimensional  model the  numbers  $(\delta \xi)_{\pm}$
evidently correspond to the  light-cone coordinates $z$ and $\zeta$ in
Section  \ref{lundmodel}. The  indices $\pm$  can be  given  a further
meaning  for the  general directrix.  According to  Eq. (\ref{xpoint})
there is  a left-moving  and a right-moving  wave spanning  the string
surface and  according to Eq.   (\ref{Aperiod}) the fixed  phase parts
(from  now on  we will  call  them ``grains'')  ``stream'' across  and
bounce back  (thereby turning from  left- to right-movers) at  the $q$
and $\bar{q}$ ends.

\begin{figure}
\begin{center}
\includegraphics{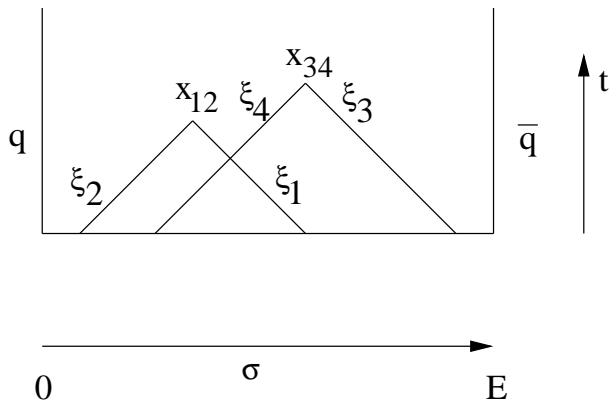}
\end{center}
\caption{The   break-up  points  and   the  corresponding   right  and
left-moving  grains in  the  ($\sigma$,t)-plane, as  described in  the
text.} \label{sigmat}
\end{figure}

We   may    describe   the   situation    in   a   $(\sigma,t)$-plane,
cf.  Fig.\ref{sigmat}, with  the right-and  left moving  grains moving
along fixed  ``light-cone lines''.  The whole string  is, at  the time
$t=0$, gathered into a single point at the origin. The boundary values
at $\sigma=0$ and $\sigma=E_{tot}$ are the directrix ${\cal A}(t)$ and
the   orbit   of  the   $\bar{q}$,   i.e.   ${\cal   A}_{\bar{q}}(t)$,
respectively.   A breakup  point $x\equiv  x_{12}$ is  reached  by the
meeting   of   the  left-moving   grain   indexed   $\xi_1$  and   the
``right-mover'' $\xi_2$ and the  breakup at $x_{34}$ by the left-mover
$\xi_3$  and   right-mover  $\xi_4$.  The   energy-momentum  $p_{34}$,
streaming into  the hadron produced between $x_{12}$  and $x_{34}$, is
evidently  $ \frac{1}{2}({\cal  A}(\xi_3)-{\cal  A}(\xi_1))$ from  the
left  and  $ \frac{1}{2}({\cal  A}(\xi_2)-{\cal  A}(\xi_4))$ from  the
right. If we  use $\pm$ as indices for the  left- and the right-movers
respectively we obtain a useful parametrisation.

With  regard to  surface areas  there is  an obvious  (scalar) surface
element  defined by the  surface spanned  by the  (incremental) grains
coming from the left and the right.
\begin{eqnarray}
\label{surfele}
dA \propto d{\cal A}_+ \cdot d{\cal A}_-
\end{eqnarray}
 Using this  we may easily  calculate the area ``below  the vertices''
$x_{12}$  and  $x_{34}$ (cf.  Fig.\ref{sigmat})  as  they are  defined
above.  They  are  the  correspondence  to the  squared  proper  times
$\Gamma$    considered   in    Section    \ref{lundmodel}   (in    the
Fig.~\ref{1particleb} they  would correspond  to the areas  $I+IV$ and
$III+IV$).
\begin{eqnarray}
\label{proptime}
\Gamma_{12}=x_{12}^2 & \mbox{and} & \Gamma_{34}=x_{34}^2
\end{eqnarray}
So, for the  vertices shown in Fig.\ref{sigmat}, these  areas are just
the squared  proper times of the  vertices. But we note  that they can
not be expressed  as the products of light-cone  components or even in
terms  of products  of  sets  of left-  and  right-movers solely,  for
general multigluon strings.

The area  below the vertices (if  we neglect the squared  mass term in
the Area  Law, which  can evidently be  included in  the normalisation
constants $N$ in Eq. (\ref{arealaw})) can be expressed as
\begin{eqnarray}
\label{3area}
A_{1234}= x_{12}^2+x_{34}^2-x_{14}^2
\end{eqnarray}
(this  is  the  correspondence   to  the  area  called  $I+III+IV$  in
connection  with  Eq.   (\ref{nonlinear})).  We  note  that  the  area
corresponding  to  $IV$, i.e.  $x_{14}^2$  in  Eq. (\ref{3area}),  may
vanish  in  this  case,  i.e.  the  left-mover  indexed  $1$  and  the
right-mover indexed $4$ may meet ``before'' $t=0$. This means that the
condition  in Eq  (\ref{nonlinear}) can  not  be met  for the  general
situation.

We may  nevertheless include these  considerations to devise  an exact
version  of the  Area  Law in  Eq.  (\ref{arealaw}). We  write, in  an
obvious way, the differentials  $d^2p \rightarrow (dp_{+} \cdot dp_-)$
and include the mass-shell  condition by a $\delta$-distribution as in
Eq. (\ref{arealaw}).   The result is,  however, very complex  to treat
and for our purposes there are three shortcomings:

\begin{itemize}

\item[1] If we  treat the process in an iterative  way, at every step,
just as we mentioned above, there are two numbers $(\delta \xi)_{\pm}$
to be  solved for.  One is  fixed by the mass-shell  condition and the
other  can be  fixed by  the  area exponential;  but the  distribution
functions  are very  complex (cf.  the remarks  under $2$  below).  In
principle everything can be done by Monte Carlo simulation methods but
even with very fast computers  and the best possible computer routines
this model  will always be much  slower than a process  where there is
only one number to be solved for, at every step.

\item[2] Due to the complexity of  the formulae there is no way we can
obtain  useful  analytical  tools   to  study  the  behaviour  of  the
correspondences   to    the   transition   operators    as   in   e.g.
Eq. (\ref{diagonal}).  This also implies  that there is no way that we
can  define the  total integral  and  sum over  the produced  hadronic
states  (as   we  could  do  for  the   $(1+1)$-dimensional  model  in
Eqs.  (\ref{Req}) and (\ref{aint}))  so as  to be  able to  obtain the
correct weighting in every step of an inclusive cascade.

\item[3] Although this implementation of  the Area Law is an ``exact''
procedure,  fulfilling  at every  step  the  mass-shell condition  and
taking  the  ``true''  area   below  the  vertices  into  account,  it
nevertheless starts out from the assumption that the partonic state is
so  well defined  that there  are  reasons to  implement an  ``exact''
theoretical fragmentation procedure on it.  This is in general not the
case, i.e. the  partonic state defined by perturbation  theory is only
well-defined  down to  some virtuality  cut-off as  we  have discussed
before.

\end{itemize}

We use quotation-marks on the word ``exact'' in order to indicate that
the use of a ``frozen  surface'' is in itself an assumption introduced
to mend  the problems obtained from  the use of  classical physics and
neglect of  the subsequent angular momentum production.   In the third
item above,  we are coming back  to the statement at  the beginning of
the  section,   that  infrared  stability  should   imply  that  minor
deformations of  the boundary  (in this case  changes in  the partonic
state as  defined by the directrix ${\cal  A}$ up to the  order of the
hadronic mass scales) should not  be noticable in the results.  In the
next section we will define a  procedure which will mend all the three
shortcomings mentioned above.

\section{The Lund String Fragmentation as a process along the directrix}
\label{method2}

In this  section, we  will consider a  fragmentation process  which is
defined  along  the directrix  and  does  not  suffer from  the  three
problems considered at the  end of Section \ref{genprocess}.  In order
to exhibit the  idea we will start with  the $(1+1)$-dimensional model
and rewrite it  in a useful way.  After that we will extend  it to the
general case. In particular, we  will allow modifications of the order
of the  hadronic mass-scale  in the directrix  in accordance  with the
discussion under point $3$ above.

\subsection{The directrix process for the $(1+1)$-dimensional case}
\label{method1+1}

In this case the directrix  contains only two directions, given by the
$\bar{q}$  energy-momentum  vector  (to  be  called  ${\cal  A}_+$  in
accordance with  the notation introduced  in Section \ref{genprocess})
and  the $q$  energy-momentum (${\cal  A}_-$). A  vertex  point $x_j$,
obtained  after  the production  of  $j$  hadrons  from the  $q$-side,
$p_1,\ldots,p_j$ is then described (with respect to the origin) by
\begin{eqnarray}
\label{xdefine}
x_j=\frac{1}{2}({\cal A}_{+j}+{\cal A}_{-j})
\end{eqnarray}
We also know that
\begin{eqnarray}
\sum_1^j p_{\ell}=\frac{1}{2}({\cal A}_{+j}-{\cal A}_{-j})
\end{eqnarray}
Using  the symmetry  of a  directrix  passing through  a single  point
(Eq. (\ref{Asym})) we may find another point on the directrix with the
property
\begin{eqnarray}
{\cal  A}_{-j}   \equiv{\cal  A}_-(\xi_j)=-{\cal  A}_-(-\xi_j)  \equiv
-{\cal B}_{-j}
\end{eqnarray}
We  will from  now on  drop the  indices ${\pm}$  on ${\cal  A}_+$ and
${\cal B}_-$  but we  note that  they do describe  points on  the same
directrix  in accordance with  the left-  and right-mover  notation in
Section  \ref{genprocess}.  While ${\cal  A}_-$ goes  ``backward'' for
increasing $j$-values, ${\cal B}$ follows the $q$-direction.

\begin{figure}
\begin{center}
\includegraphics{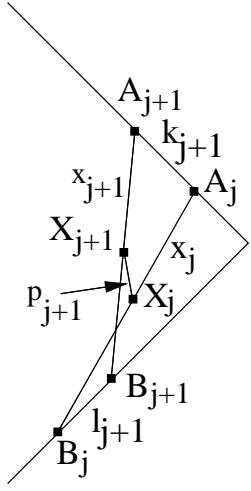}
\end{center}
\caption{A  description   of  the  fragmentation   process  along  the
directrix with notations according to the text.} \label{prodir}
\end{figure}

We may now  consider the hadron energy momenta to  define a curve from
the origin  ``along the directrix'' such  that after $j$  steps it has
reached the point
\begin{eqnarray}
\label{hadroncurve}
\sum_1^j p_{\ell} \equiv X_j = \frac{1}{2}({\cal A}_j + {\cal B}_j)
\end{eqnarray}
while the difference  between the point ${\cal A}_j$  on the directrix
and $X_j$ is given by  $x_j$ in Eq. (\ref{xdefine}). The production of
a new particle  $p_{j+1}$ then corresponds to choosing  two new points
$X_{j+1}$ and  ${\cal A}_{j+1}$ (along  the directrix) such  that, cf.
Fig.\ref{prodir}
\begin{eqnarray}
\label{stepdir}
X_{j+1}-X_{j}=p_{j+1}  &  \mbox{and}  &  {\cal  A}_{j+1}-{\cal  A}_{j}
\equiv k_{j+1}
\end{eqnarray}
We also  obtain a  new ``vertex'' vector  $x_{j+1}$ by the  identity (
cf. Fig.\ref{prodir}):
\begin{eqnarray}
\label{krelation}
p_{j+1}+x_{j+1}=x_j + k_{j+1}
\end{eqnarray}

\noindent  In  this  way  the  vertex  vector  fulfils  $x_{j+1}={\cal
A}_{j+1}-X_{j+1}$ just as $x_j={\cal A}_j- X_j$. We have then arranged
it so that the hadrons are produced along a curve, the $X$-curve, from
the origin  and the vertex vectors  are the connectors  for this curve
going from the produced particle to the directrix.  Before we consider
the Area Law in this situation we note the symmetry between a reversed
process and the process described above, i.e. when we go from $X_j$ to
$X_{j+1}$ producing  $p_{j+1}$ by the use  of a part  $k_{j+1}$ of the
directrix along ${\cal A}$.

\begin{figure}
\begin{center}
\includegraphics{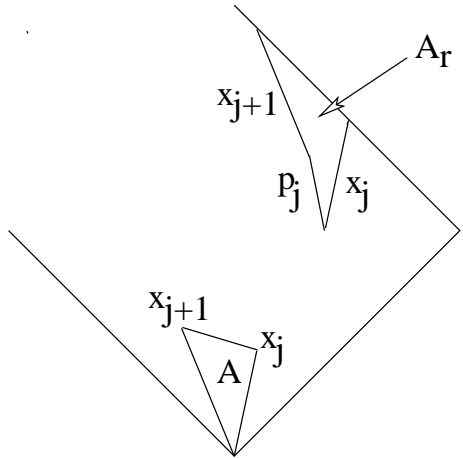}
\end{center}
\caption{The figure shows both  the production as described in Section
\ref{lundmodel}  and   the  $k$-process  described  in   the  text.  }
\label{process}
\end{figure}

To see the  reverse process we note that the vector  $x_j$ can just as
well be  reached by taking the  difference between the  point $X_j$ on
the hadron curve, Eq.  (\ref{hadroncurve}), and ``the backward point''
on the directrix ${\cal B}_j$.
\begin{eqnarray}
x_j =\frac{1}{2}({\cal A}_j  - {\cal B}_j) = {\cal A}_j -  X_j = X_j -
{\cal B}_j
\end{eqnarray}
Using this we could evidently  consider the production of the particle
$p_{j+1}$ as  a step from  ${\cal B}_j$ to ${\cal  B}_{j+1}={\cal B}_j
+\ell_j$ (cf. Eq. (\ref{stepdir})) such that we have in correspondence
to Eq. (\ref{krelation}) and Fig.\ref{prodir}
\begin{eqnarray}
\label{lrelation}
p_{j+1}+x_{j}=x_{j+1}+\ell_{j+1}
\end{eqnarray}
In  order  to  formalise  the  determination of  the  particle  energy
momentum $p$,  we may then  in ``the $k$-process'' (along  ${\cal A}$)
assume that  we know  the starting vertex  vector $x$, connected  to a
point ${\cal A}_P$.   We may then choose a piece  $k$ along ${\cal A}$
(of  a size to  be determined)  and then  define the  other light-cone
direction in the plane determined by $(x,k)$ by
\begin{eqnarray}
\label{lhat}
\hat{\ell}= x - k \frac{x^2}{2xk}
\end{eqnarray}
The vector $p$ will be described in terms of $(k,\hat{\ell})$ as
\begin{eqnarray}
\label{pdefine}
p=z\hat{\ell} + \frac{k}{2}=zx +\frac{k}{2}(1-\frac{zx^2}{xk})
\end{eqnarray}
with the requirement that the particle should be on the mass-shell
\begin{eqnarray}
\label{zdefine}
p^2=m^2 = z k x &\mbox{i.e.} & kx=\frac{m^2}{z}
\end{eqnarray}
From   Eq.  (\ref{krelation})   we  obtain   the  new   vertex  vector
$x^{\prime}$ by
\begin{eqnarray}
\label{xprime}
x^{\prime}=(1-z)   x   +  \frac{k}{2}(1+\frac{zx^2}{xk})   \nonumber\\
(x^{\prime})^2 = (1-z)(x^2+ x k)=(1-z)(x^2+\frac{m^2}{z})
\end{eqnarray}
In  Fig.\ref{process} we  show  both the  production  as described  in
Section  \ref{lundmodel}  and  the  $k$-process  described  above.  In
particular we note  the two areas exhibited. It  is straightforward to
prove that
\begin{eqnarray}
\label{twoareas}
A= A_r +\Gamma^{\prime}-\Gamma
\end{eqnarray}

\noindent with  the obvious definition of e.g.  $\Gamma=x^2$. The Area
Law in  Eq. (\ref{arealaw}) will be  fulfilled both by the  use of the
area in Fig.\ref{area}  and the area in between  the directrix and the
hadronic curve (the $X$-curve) if  we choose the variable $z$ from the
fragmentation  function  in  Eq.  (\ref{aint})  and  apply  it  as  in
Eqs.   (\ref{pdefine})-(\ref{xprime}).   This   is   so  because   the
difference  between  the areas  $A$  and  $A_r$  for the  single  step
exhibited above will vanish when we consider the whole process.

It is also obvious that we may define an $\ell$-process similar to the
$k$-process    we    have    discussed    above.   We    just    write
$z\hat{\ell}=\half\ell$ and  introduce the variable  $\zeta$ such that
$\half k=\zeta \hat{k}$ with $\zeta$ chosen such that $m^2/\zeta= \ell
x^{\prime}$. Actually  we obtain the  same process (although  ``in the
opposite order'')  under the assumption that we  start at $x^{\prime}$
and chose $\ell$  along the ${\cal B}$-part of  the directrix with the
variable $\zeta$ in accordance with Eq. (\ref{zeta}).  In this way the
``backward'' variable $\zeta$ evidently obeys the same distribution as
the ``forward'' variable $z$ and the Area Law is fulfilled.

We have  reached a situation  where a particle production  step starts
from a knowledge of a  vector $x$ connected to a light-cone-direction.
Then we choose  a light-like vector $k$ such  that Eq. (\ref{zdefine})
is  fulfilled   with  a  $z$-value  stochastically   chosen  from  the
fragmentation function  in Eq.  (\ref{aint}). After  that we construct
the particle  energy-momentum and a new  vector $x^{\prime}$ according
to Eqs. (\ref{pdefine}) and  (\ref{xprime}). We may start out choosing
the  ``first''  $x$-vector  ($x_0$)  equal  to  the  $q$  (light-cone)
energy-momentum and then the $(1+1)$-dimensional model is defined.

\subsection{The directrix process around a gluon corner}
\label{gluoncorner}
For  a multi-gluon directrix  however, we  will reach  situations when
there  are ``corners'',  i.e.  where the  directrix changes  direction
between two  colour-connected partons. Thus besides  the vertex vector
$x$, there will  be a remaining light-cone vector  (to the corner) $c$
and a chosen number $z$ such that
\begin{eqnarray}
\label{cdefine}
cx <\frac{m^2}{z}
\end{eqnarray}
We will now  consider ways to pass over such  a corner.  Assuming that
the directrix continues on the other  side of the corner we may choose
a  light-cone vector $k$  which fulfils  Eq. (\ref{zdefine})  (we note
that according to Eq. (\ref{cdefine}) $k$ can not be chosen along $c$)
and another light-cone vector $c^{\prime}$ such that
\begin{eqnarray}
\label{cprime}
k+c^{\prime}=c+ \hat{Q} \equiv Q
\end{eqnarray}
with $\hat{Q}$,  a vector pointing from  the corner to a  new point on
the directrix,  and $Q$  describing that point  from the point  on the
directrix where the vertex vector $x$ is connected.

We note that with this definition the suggested way to pass the corner
corresponds  to an  exchange of  one ``dipole'',  $c +\hat{Q}$  in the
process  for another $k  +c^{\prime}$, i.e.  to make  a detour  onto a
``new'' directrix which is ``close''  to the original one.  The choice
of the new dipole corresponds to  a definition of what we will mean by
the allowed ``minor deviations'' of the original perturbative state.

In order to investigate what  can be allowed under these conditions we
will exhibit a  few possible choices to fulfil  Eq. (\ref{cprime}). We
will then find  that all possible choices are not  allowed if we would
like  to  obtain  reasonable   results  for  the  final  state  hadron
distributions.  A first set of possibilities is to choose $c^{\prime}$
as a new light-like connector  that will bring us ``closer'' (than the
vector $c$)  to the  original directrix in  the next  production step.
One  may then  hope that  after another  step in  the process  the new
connector becomes even ``smaller'' so that we are back at the original
directrix after a few steps.

\begin{itemize}

\item[C1] The  vector $k$ can  be chosen in  the plane spanned  by the
vertex vector  $x$ and the  directrix $Q$ (then both  $c^{\prime}$ and
$k$ are fully determined).

\item[C2]  A more  general choice  is to  take the  vector $k$  in the
three-space spanned by $x$, $Q$  and the remainder vector $c$. In this
case it  is necessary to provide  a second condition  to determine $k$
and  $c^{\prime}$.  In  the rest-frame  of the  vertex vector  $x$ the
space  parts  of the  vectors  $\vec{c}$  and  $\vec{\hat{Q}}$ span  a
triangle     with     the     vector    $\vec{Q}$     as     baseline,
cf.  Fig.\ref{2space}. (For  the case  exhibited we  assume  that also
$\hat{Q}$ is  light-like.  The situation can be  easily generalised to
the  situation  when  there  are  one  or  more  ``corners''  also  on
$\hat{Q}$.)  The remaining degree of  freedom will then be fixed by an
angular   condition.  One   such   condition  is   indicated  in   the
Fig.\ref{2space}. The  two triangles $(\vec{c},\vec{\hat{Q}},\vec{Q})$
and  $(\vec{c^{\prime}},\vec{k},\vec{Q})$  in  this  construction  are
chosen congruent. This is a kind of minimal choice for the size of $Q$
in this  three-space. It is not  difficult to see that  in general the
new connector will be smoothly connected to the original directrix.

\begin{figure}
\begin{center}
\includegraphics{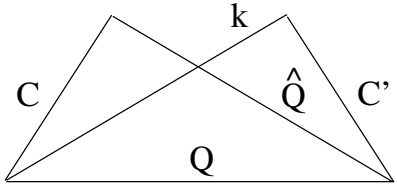}
\end{center}
\caption{A possible construction (C2)  for passing a gluon corner. The
figure shows the space parts of  the vectors, in the rest-frame of the
vertex vector $x$, as described in the text.} \label{2space}
\end{figure}

\end{itemize}

For  the case  defined by  the  condition C1  we obtain  that $k$  and
$c^{\prime}$ will  be directed along the two  light-cone directions in
the plane spanned by $(x,Q)$:
\begin{eqnarray}
\label{kcprime}
k=\frac{Q}{2}+ \frac{((Qx)Q-Q^2x)}{2\sqrt{(Qx)^2-Q^2x^2}} & \mbox{and}
& c^{\prime}=Q-k
\end{eqnarray}
(In    Fig.\ref{2space}   they   would    be   directed    along   the
$\vec{Q}$-direction   with   the   lengths   $(Q_0\pm   |\vec{Q}|)/2$,
respectively). The vector $Q$ contains one degree of freedom which can
be fixed  by multiplying  with $x$ in  Eq. (\ref{kcprime}),  and using
Eq. (\ref{zdefine}).

This means on the one hand  that the situation corresponds to using up
the  smallest possible segment  of the  directrix in  the construction
according to Eq.  (\ref{cprime}).  On the other hand  we will obtain a
very  sharp ``bend''  on the  new  directrix segment,  defined by  the
dipole $(k c^{\prime})$.  While the first result is  desirable from an
economy  point of  view, the  second implies  that we  will  break the
coherence  conditions  for the  directrix  (the  angular ordering)  as
discussed in Section \ref{cohQCD}.

The fragmentation distributions obtained with this choice are not very
encouraging.   After  we ``pass''  the  gluon  corner  the density  of
hadrons is  much higher  than the density  before the corner.  {\it We
find that even for a small gluonic excitation a break of coherence may
result in  the production  of too many  hadrons over a  large rapidity
range}.

One might hope that the situation  may be mended by a suitable angular
choice. This is so for the  situation described in C2 for a large part
of   the   events  and   for   simple   gluonic  configurations.    In
Fig.\ref{rapidity15}  we   show  a  Monte  Carlo   simulation  of  the
(inclusive)  final state  hadron rapidities  in a  state  containing a
single gluon  placed at  the centre with  a transverse momentum  of 15
GeV. Even though for most  fragmentation events the choice C2 performs
well, on the level of a few percent it does happen that the directions
of  the connecting $c^{\prime}$  and the  new vertex  $x^{\prime}$ are
such that  we obtain  a situation close  to the  one for C1.   The new
directrix (possibly after a few further steps in the process) will all
the time  come closer  to the original  one; but  at the same  time we
obtain a sharp  bend farther and farther away  from the original gluon
corner with each  step. This causes too many  particles to be produced
over a large rapidity range (cf. Fig.\ref{rapidity15}).

It  is possible to  make further  changes in  the procedure  with more
sophisticated angular choices, but we  will instead go over to another
and more successful choice, which we will call C3.
\begin{itemize}
\item[C3]A natural  choice for the  vector $c^{\prime}$ is to  make it
into the $k$-vector  for the next particle production  in the process,
i.e.    chose   a   new   stochastic  value   $z^{\prime}$   and   put
$c^{\prime}=k^{\prime}$ with
\begin{eqnarray}
\label{kprime}
k^{\prime}x^{\prime}=\frac{m^2}{z}^{\prime}
\end{eqnarray}
in  terms  of  the  new  vertex vector  $x^{\prime}$  determined  from
$(z,k,x)$  according  to  Eq.  (\ref{xprime}).   In this  way  we  are
evidently getting back to the original directrix as fast as possible.
\end{itemize}

\noindent We may  define the vector $k$ still in  the space spanned by
$(c,x,Q)$:
\begin{eqnarray}
\label{kdefini}
k=\alpha c + \beta x + \gamma Q
\end{eqnarray}
We obtain a solution for the coefficients $(\alpha,\beta,\gamma)$ from
the requirements
\begin{eqnarray}
\label{kdefini2}
k^2=0 \nonumber\\ kx = \frac{m^2}{z} & \mbox{and} & kQ=\frac{1}{2}Q^2
\end{eqnarray}
(these conditions also imply  that $k^{\prime}$ is light-like), and an
equation  for   the  vector  $Q$   along  the  directrix   using  Eqs.
(\ref{kdefini2}) and
\begin{eqnarray}
\label{Qdefini}
\frac{m^2}{z^{\prime}}       =k^{\prime}x^{\prime}=       (Q-k)((1-z)x
+\frac{k}{2}(1+ \frac{zx^2}{kx}))
\end{eqnarray}
In  the  fragmentation  distribution   there  is  a  factor  $(1-z)^a$
(stemming  from the  density  of states  according  to the  discussion
around  Eq. (10)).   This  implies that  we  may not  use  up all  the
energy-momentum  along one of  the light-cone  directions in  a single
step. Small values of $z$  correspond to $\zeta \simeq 1$ according to
Eq. (11),  i.e. to using up  almost all the  energy-momentum along the
other light-cone  direction in a  step. Such values are  suppressed by
the area  exponential in the distribution. Nevertheless  it may happen
that we obtain a very small $z$-value ($z \leq 0.1$), corresponding to
a large  (but not  infinitely large) area.  Sometimes for  the general
multi-gluon  directrix  however,  there  might be  no  possibility  to
accomodate such a large area due to energy-momentum conservation. Then
the  choice C3  can  not be  done.  Instead of  introducing a  general
projection of  the two-particle distributions  for all such  cases, in
the Monte Carlo simulation we have  chosen to go back to the beginning
and start a new fragmentation  event. Such situations, when we can not
use  the  choice C3,  are  infrequent  enough  to let  this  ``restart
strategy'' be usable.

In  order to  test  the influence  of  this feature  on  the model  we
investigate  the influence  on  the single  particle and  two-particle
distributions  below  in  Section   \ref{results}  and  we  find  them
negligible.  In order  to show  that  our changes  in the  directrices
actually  are of  the  hadronic mass  scale  we have  defined a  local
``bending parameter'' closely related to the $k_{\perp}$ variable used
in the ordering of the Lund Dipole Model, cf. Eq. (\ref{lambda1})
\begin{eqnarray}
\label{kperp2def}
k^2_{\perp 2}= \frac{2 (xk_1)(k_1k_2)}{x(k_1+k_2)+k_1k_2}
\end{eqnarray}
From the results  in Section \ref{results} we conclude  that we are at
most making  local modifications  on the order  or below  the hadronic
scale in our directrices.

\subsection{A Differential Process and its Relationship to the Generalised
Rapidity}
\label{diffcurve}

There is  a direct  connection between a  differential version  of our
hadronisation process and the ${\cal X}$-curve that was referred to in
the  Introduction,  \cite{BAGGBSII}.  In  order  to  investigate  this
process  we  consider the  limiting  situation  for  a vanishing  mass
parameter.  Then the  distribution function  will obviously  develop a
pole for $z  \rightarrow 0$. We will assume that  the model is defined
by the incremental step size $dz$ with the ratio $m/dz \rightarrow m_0
$.  The  corresponding incremental $k$  vector will be  called $d{\cal
A}$  and it  will fulfil  the mass-shell  condition (we  will  use the
notation $q_P$ instead of $x$ for the vertex vector in connection with
the differential process)
\begin{eqnarray}
\label{dzdef}
q_Pd{\cal A} = \frac{m^2}{dz} \rightarrow dz m_0^2
\end{eqnarray}
From  the model  formulae for  the change  in $q_P$  and  the particle
energy-momentum  $p$  (Eqs.  (\ref{pdefine})  and  (\ref{xprime}))  we
obtain  the following differential  equations defining  a curve  to be
called the ${\cal P}$-curve
\begin{eqnarray}
\label{pcurve}
d{\cal  P}  =  dz  q_P  +  \frac{d{\cal  A}}{2}(1-\frac{q_P^2}{m_0^2})
\nonumber\\      dq_P      =      -dz     q_P      +      \frac{d{\cal
A}}{2}(1+\frac{q_P^2}{m_0^2})
\end{eqnarray}
Firstly  we  note  that from  the  sum  and  differences of  the  Eqs.
(\ref{pcurve}) we obtain
\begin{eqnarray}
\label{Pplusq}
{\cal P} + q_P = {\cal A} \nonumber\\ {\cal P} - q_P = {\cal L}
\end{eqnarray}
where  the vector  ${\cal L}$  has a  light-like tangent  just  as the
directrix ${\cal A}$:
\begin{eqnarray}
\label{calLdef}
d{\cal L}=2 dz q_P -d{\cal A}\frac{q_P^2}{m_0^2}
\end{eqnarray}
In  this way  the ${\cal  P}$-curve goes  in between  two  curves with
everywhere light-like  tangents and the vector $q_P$  connects to both
of them.

The vector $q_P$ is time-like and quickly approaches the length $m_0$.
To see that we multiply the second line of Eq. (\ref{pcurve}) by $q_P$
and obtain
\begin{eqnarray}
\label{qPTlength1}
dq_P^2=dz(-q_P^2 +m_0^2)
\end{eqnarray}
This  means (remembering  that  $dz=q_Pd{\cal A}/m_0^2$)  that we  may
write:
\begin{eqnarray}
\label{qPTlength2}
q_P^2    =     m_0^2    (1-T_P^{-1})    &     \mbox{with}    &    T_P=
\exp(\int(\frac{q_Pd{\cal A}}{m_0^2}))
\end{eqnarray}
We have  then assumed  that $q_P=0$  and $T_P=1$ at  the start  of the
process.  We note  that the integrand in the exponent  for $T_P$ is an
area, more  precisely the area  betweeen the directrix and  the ${\cal
P}$-curve.   This  is very  similar  to  the  results we  obtained  in
\cite{BAGGBSII} for  the generalised rapidity and we  will now briefly
connect to these results.

\subsection{The ${\cal X}$-curve and its Properties}
\label{Xcurve}

The distribution in Eq.  (\ref{arealaw}) contains two terms, the phase
space and the exponential area suppression. In order to obtain a large
probability it is  necessary for a given total  energy-momentum on the
one hand to make many particles on the other hand to make them in such
a way that the area is small. The obvious compromise is that the decay
region is around a typical  hyperbola with an average squared distance
to the origin $<\Gamma>\equiv  \Gamma_0$.  The length of the hyperbola
is proportional  to the available  rapidity range for the  final state
particles, i.e.  $\Delta y =  \log(s/\Gamma_0)$ with s the squared cms
energy.

\begin{figure}
\begin{center}
\includegraphics{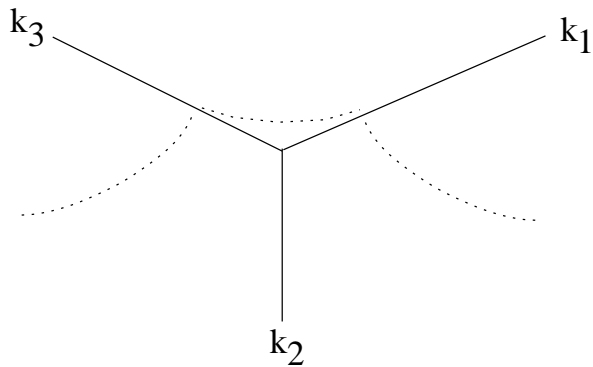}
\end{center}
\caption{A string  with a $q(k_1)$, $\bar{q}(k_3)$ and  a single gluon
excitation $(k_2)$.  The figure  shows the connected region around the
gluon ``tip''. } \label{onegluon}
\end{figure}

In  the Lund Model  interpretation for  a string  with a  single gluon
excitation, there will be two parts of the string; one spanned between
the $q$ and the $g$ and one between the $g$ and the $\bar{q}$. Each of
them should  break up  in a  similar way as  the single  string region
described by  Eq. (\ref{arealaw}).  Besides  that there will be  a few
particles produced  in the connected region around  the gluon ``tip'',
cf.  Fig.\ref{onegluon}.  If the  energy-momenta  of  the partons  are
$k_j$,  $j=1,2,3$ (with indices  $1$ ($3$)  for the  $q$ ($\bar{q}$)),
then there  will be two  hyperbolic angular ranges,  $(\Delta y)_{12}$
and $(\Delta y)_{23}$.  The sum of these ranges is
\begin{eqnarray}
\label{lambda1}
\lambda=(\Delta          y)_{12}+         (\Delta         y)_{23}=\log
(\frac{s_{12}}{2\Gamma_0})+ \log(\frac{s_{23}}{2\Gamma_0})\nonumber \\
= \log(\frac{s}{\Gamma_0}) + \log(\frac{s_{12}s_{23}}{4\Gamma_0 s})
\end{eqnarray}
Here $s_{j\ell}= (k_j+k_{\ell})^2$  and $s = s_{12}+s_{23}+s_{13}$ and
the  factors  $2$  are  introduced  because only  half  of  the  gluon
energy-momentum goes into each string region.

The quantity $k_{\perp}^2 \equiv  s_{12}s_{23}/s$ is a convenient (and
Lorentz invariant)  approximation for  the transverse momentum  of the
emitted  gluon. From Eq.  (\ref{lambda1}) we  conclude that  after the
emission  of a  single gluon  the phase  space is  increased  from the
single  hyperbola  result  above  by  an  amount  corresponding  to  a
``sticking-out tip'' of  length given by the logarithm  of the emitted
transverse momentum.  In conventional notions this  corresponds to the
``anomalous  dimensions''  of  QCD,  i.e.  the  emission  of  a  gluon
increases the  region of colour flow  inside which more  gluons can be
emitted and hadronisation can take place. The whole scenario is easily
visualised  and used  in the  Lund  Dipole Model,  \cite{GG}, and  the
corresponding Monte Carlo simulation program ARIADNE, \cite{LL}.

It is straightforward to see that  if there are many gluons then there
is a  corresponding quantity,  a generalised rapidity  $\lambda \simeq
\log (\prod  s_{j j+1})$ stemming from the  hyperbolae spanned between
the colour-connected gluons. But we  note that this is not an infrared
stable definition. We will now provide a convenient generalisation.

A closer examination  of the region around the tip  of a gluon reveals
that there is  a correction corresponding to a  connected hyperbola in
the region  $(k_1,k_3)$ between the  ``endpoint'' of the  hyperbola in
the region  spanned between $(k_1,k_2/2)$ and the  one spanned between
$(k_2/2,k_3)$, cf. Fig.\ref{onegluon}.  In  formulae we obtain for the
average hyperbolae
\begin{eqnarray}
\label{lambda2}
(\alpha_1k_1+\frac{1}{2}\beta_1   k_2)^2=\Gamma_0   &   \mbox{and}   &
(\gamma_3  k_3+  \frac{1}{2}\beta_3   k_2)^2  =\Gamma_0  \nonumber  \\
(\alpha_2 k_1 + \gamma_2 k_3 +\frac{1}{2}k_2)^2 =\Gamma_0
\end{eqnarray}
with    the   ranges    $1   \geq    \alpha_1\geq   2\Gamma_0/s_{12}$,
$2\Gamma_0/s_{12}  \geq  \alpha_2   \geq  0$,  $2\Gamma_0/s_{12}  \leq
\beta_1\leq 1$  and similarly for  the other variables. The  length of
the two hyperbolae in the segments $(k_1,k_2/2)$ and $(k_2/2,k_3)$ are
then given by Eq. (\ref{lambda1})  but the third hyperbola provides an
extra contribution (in the  appropriate limit $s_{13} \simeq s$) equal
to $\log(1 + 4\Gamma_0  s/s_{12}s_{23})$. Then the total (generalised)
rapidity length becomes
\begin{eqnarray}
\label{lambda3}
\lambda_{123}           =           \log(\frac{s}{\Gamma_0}          +
\frac{s_{12}s_{23}}{(2\Gamma_0)^2})
\end{eqnarray}
This is evidently a nice interpolation between the situations with and
without  a gluon  on the  string  and it  is also  an infrared  stable
definition of  the notion of  rapidity length. Eq.  (\ref{lambda3}) is
noted in \cite{BAGGBSII} and led  us to introduce a functional defined
on a multigluon string directrix.

We may firstly define a set of connected integrals, \cite{BAGGBSII}:
\begin{eqnarray}
\label{In}
I_n= \int ds_{01}ds_{12}\cdots ds_{nE}
\end{eqnarray}
with the  easily understood notation (cf.  Eq. (\ref{momentum})) $s_{j
j+1}=({\cal A}(\xi_j)-{\cal A}(\xi_{j+1}))^2$, i.e. it is proportional
to the squared  mass between the points $\xi_j$  and $\xi_{j+1}$ along
the  directrix.  By  performing  the  integrals, we  obtain  that  the
argument in the  logarithm in Eq. (\ref{lambda3}) is  given by the sum
$I_1/\Gamma_0 +  I_2/(2\Gamma_0)^2$ and that we may  in general define
the functional $T$ by :
\begin{eqnarray}
\label{T}
T=1+ \sum_{n=1}^{\infty}\frac{I_n}{(2m_0^2)^n}
\end{eqnarray}
as a suitable generalisation for any string state. For a finite number
of partons $N$ the terms with $n > N$ will all vanish and we also note
that the highest degree term will always have the generic form :
\begin{eqnarray}
\label{Tnterm}
2     \frac{s_{12}}{4m_0^2}\frac{s_{23}}{4m_0^2}\cdots    \frac{s_{N-1
N}}{4m_0^2}
\end{eqnarray}
We also note that for a  finite total energy $E$ the contributions for
very large  degrees will  become smaller and  smaller compared  to the
scale $m_0$.

In order  to study the  functional $T$ it  is suitable to  introduce a
varying value $\xi$  instead of the total energy  $E$ in the connected
integrals. It is then evident that the functional $T(\xi)$ will fulfil
the integral equation :
\begin{eqnarray}
\label{Tint}
T(\xi)=     1     +    \int_0^{\xi}\frac{ds(\xi,\xi^{\prime})}{2m_0^2}
T(\xi^{\prime})
\end{eqnarray}
We will also introduce  the vector-valued function $q_T(\xi)$ together
with $T$ so that we have
\begin{eqnarray}
\label{qTint}
q_{T     \mu}(\xi)=\frac{\int_0^{\xi}d{\cal     A}_{\mu}(\xi^{\prime})
T(\xi^{\prime})}{T(\xi)}\nonumber\\                           T(\xi)=1+
\int_0^{\xi}\frac{q_T(\xi^{\prime})d{\cal      A}(\xi^{\prime})}{m_0^2}
T(\xi^{\prime})
\end{eqnarray}
By differentiation and integration we obtain the results
\begin{eqnarray}
\label{Tarea}
T=                       \exp(\int_0^{\xi}\frac{q_T(\xi^{\prime})d{\cal
A}(\xi^{\prime})}{m_0^2})    \equiv   \exp(\lambda(\xi))   \nonumber\\
q_T^2(\xi)=m_0^2(1-T^{-2}(\xi))
\end{eqnarray}
The  generalised  rapidity  $\lambda$  corresponds to  the  result  in
Eq.  (\ref{lambda3})  for  the  simple  case described  above  and  it
provides   an   infrared   stable   definition  for   any   multigluon
state. Further the vector $q_T$ is time-like and will quickly approach
the finite length $m_0$.

\subsection{The Correspondence between the ${\cal X}$-curve and
the ${\cal P}$-~curve}
\label{calPcalX}

The similarity between these results  and the results obtained for the
${\cal P}$-curve in Eqs. (\ref{qPTlength1}) and (\ref{qPTlength2}) are
obvious (besides the power in  $T^{-1}$) and there is a correspondence
to   the   Eqs.    (\ref{pcurve})   and  (\ref{Pplusq})   also.    The
interpretation   of   the  ${\cal   X}$-curve   (as   worked  out   in
\cite{BAGGBSII}, cf.  also \cite{BA}) is that there is a vector valued
function ${\cal X}_{\mu}(\lambda)$  conveniently labelled by $\lambda$
such that
\begin{eqnarray}
\label{calXdiff}
{\cal       X}+q_T={\cal       A}\nonumber       \\       \frac{d{\cal
X}}{d\lambda}=q_T\nonumber\\  \frac{dq_T}{d\lambda}=-q_T +\frac{d{\cal
A}}{d\lambda}
\end{eqnarray}
i.e. the  vector $q_T$ is the  tangent to the curve  defined by ${\cal
X}$  such  that  it reaches  to  the  directrix.  There is  no  direct
correspondence  to the ${\cal  L}$-curve (unless  the vectors  $q$ has
reached its asymptotic length $m_0$).

\begin{figure}
\begin{center}
\includegraphics{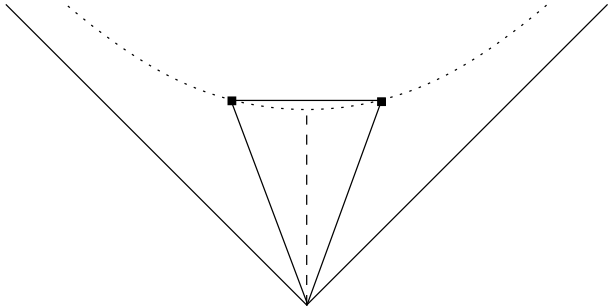}
\end{center}
\caption{Two vertices  placed symmetrically (around  the rapidity y=0)
at $m_0(\cosh(\delta  y/2), \sinh(\delta y/2))$  and $m_0(\cosh(\delta
y/2),  -\sinh(\delta  y/2))$   respectively,  with  a  distance  $2m_0
\sinh(\delta y/2))$.  } \label{stephyp}
\end{figure}

It is useful to calculate  the results from the differential equations
for the case when there is  a finite length light-like vector $k_j$ in
the  directrix. By  direct integration  we find  that if  we  have the
vector   $q_{Tj}$   then   ``after''   application   of   the   parton
energy-momentum $k_j$ we obtain the vector $q_{T j+1}$ and will take a
step along the ${\cal X}$-curve equal to $\delta {\cal X}_j$
\begin{eqnarray}
\label{intqTcalX}
q_{T  j+1}=\gamma_j q_{T  j}  +\frac{(1+\gamma_j)}{2} k_j  \nonumber\\
\delta{\cal   X}_j=(q_{T   j}+\frac{1}{2}  k_j)(1-\gamma_j)\nonumber\\
\gamma_j=\frac{1}{1+\frac{q_{Tj}k_j}{m_0^2}}
\end{eqnarray}
(we  also  note  that the  products  of  the  $\gamma_j$ is  equal  to
$T^{-1}$).  The  correspondence  to   these  results  for  the  ${\cal
P}$-curve are
\begin{eqnarray}
\label{intqPcalP}
q_P^{\prime}=  \gamma  q_P +\frac{(1+\frac{q_P^2\gamma}{m_0^2})}{2}  k
\nonumber     \\     \delta      {\cal     P}=     (1-\gamma)q_P     +
\frac{(1-\frac{q_P^2\gamma}{m_0^2})}{2} k \nonumber \\ \delta {\cal L}
\equiv \ell=  2 (1-\gamma)q_P -  \frac{q^2_P \gamma}{m_0^2}k \nonumber
\\    \gamma    =\frac{1}{1+\frac{q_Pk}{m_0^2}}    &   \mbox{and}    &
(T_P)^{-1}=\prod \gamma_j
\end{eqnarray}
This means that while the length  of a step along the ${\cal X}$-curve
depends upon  $q_T^2$ the corresponding  step length along  the ${\cal
P}$-curve is
\begin{eqnarray}
\label{masscond}
(\delta {\cal P})^2\equiv M^2_j = \frac{(1-\gamma_j)^2m_0^2}{\gamma_j}
\end{eqnarray}
This  has a  very  simple meaning  for  a step  length  $M_j$ along  a
hyperbola  with the  parameter $m_0$,  cf. Fig.\ref{stephyp}.   If the
vertices  are placed symmetrically  around the  rapidity $y=0$  at the
positions  $m_0(\cosh(\delta y/2),  \pm \sinh(\delta  y/2))$  then the
step  length is evidently  $2m_0 \sinh(\delta  y/2))$ which  should be
compared to $M_j$.  Then if we square them and define $\gamma_j \equiv
(1-z_j)=\exp(-\delta y)$ the relationship in Eq. (\ref{masscond}) will
ensue.

The identification of $\gamma$ and  $(1-z)$ is actually a very general
feature  of  the process(we  will  come back  to  this  aspect in  the
future).   At   this  point   we  only  note   that  if   the  lengths
(proper-times)    of   the   two    adjacent   vertices    are   equal
$\Gamma=\Gamma^{\prime}=m_0^2\equiv      <\Gamma>$      then      from
Eq. (\ref{Gammarel}) we obtain
\begin{eqnarray}
\label{equalgamma}
m_0^2= \frac{(1-z)m^2}{z^2}
\end{eqnarray}
which  evidently coincides  with the  result in  Eq. (\ref{masscond}).
Therefore, for this particular value of $z$, the fragmentation process
will all  the time  proceed along the  connected hyperbolae. It  is in
this way  that the ${\cal P}$-curve  can be considered  the average of
the hadronic $X$-curve.

In order to  provide a precise relationship between  the vectors $q_P$
 and $q_T$ as well as the functionals  $T_P$ and $T$ we make use of an
 interesting relationship for the ${\cal X}$-curve which is derived in
 \cite{BAGGBSII}.   If   we  define  the   $(1+4)$-dimensional  vector
 $(Q_{\mu}\equiv Tq_{T \mu}/m_0,T)$ (which has a length in the $(1+4)$
 dimensional  Minkowski   metric  equal  to   $Q^2-T^2=-1$)  then  the
 differential equation for $q_T$ can be rewritten as
\begin{eqnarray}
\label{rotations}
dT= Qd{\cal A} & \mbox{and} & dQ=TdA
\end{eqnarray}
i.e. as a group of special rotations in this space (corresponding to a
subgroup  of SO(1,4))  which are  defined by  the  incremental changes
along  the directrix  curve.  The corresponding  relationship for  the
${\cal P}$-curve is, from the Eqs. (\ref{dzdef}) and (\ref{pcurve}) :
\begin{eqnarray}
\label{rotat2}
d  (q_PT_P)=  d{\cal  A}  (T_P-\frac{1}{2})  &  \mbox{and}  &  dT_P  =
q_Pd{\cal A} T_P
\end{eqnarray}
so that the $(1+4)$-dimensional vector  $(2q_P T_P, 2T_P -1)$ has both
the  same length  and  fulfils the  same  differential equations  with
respect  to incremental  changes along  the directrix  as  $(Q,T)$. In
\cite{BAGGBSII} we characterized the different directrices by means of
the  eigenvalues of  the transfer  matrix along  the directrix  and we
found a  simple and  useful method to  relate them to  the generalised
rapidity $\lambda = \log(T)$. We will come back to this formalism in a
future publication and show the significance of the extended space.

In conclusion we  have found a differential version  of our stochastic
process which corresponds to a curve of connected hyperbolae along the
directrix function.  The  area in between the curve  and the directrix
(scaled by the  single parameter $m_0^2$) has the  interpretation of a
generalised  rapidity measure  for the  multigluon case.   It  is also
related to a  group of rotations with the  incremental steps along the
directrix  as the  generators. The  curve  may be  interpreted as  the
average hadronic  curve stemming from  the hadronisation of  the given
directrix.

\subsection{The Reverse Problem, to Find the Directrix from the
Hadronic Curve}
\label{reverse}

We  will  end this  section  by solving  the  reverse  problem to  the
hadronisation process, i.e. to exhibit to what extent we can trace the
directrix from a  knowledge of the hadronic curve,  which we will call
the $X$-curve  in accordance with  the notation introduced  in Section
\ref{method1+1}.

We  will then assume  that the  $X$-curve is  defined by  the hadronic
energy  momenta   $\{p_j\}$,  ordered   and  laid  out   according  to
rank. (Note that  this can not be done for a  real event obtained from
an experiment, since in general,  rank is not an observable.)  We will
concentrate on  the production of  the hadron $p_j$,  produced between
the vertex  vectors $x_{j-1}$ and  $x_{j}$ with the  directrix segment
$k_j$,  and  an  appropriately  distributed stochastic  number  $z_j$.
According to Eq. (\ref{krelation}), in  order to construct $k_j$ it is
sufficient to know $p_j$ and the difference vector
\begin{eqnarray}
\label{phatdef}
(x_{j}-x_{j-1})= \epsilon_j\hat{p}_j
\end{eqnarray}
It is straightforward  to solve for $\hat{p}_j$ in  terms of $p_j$ and
$x_{j-1}$
\begin{eqnarray}
\label{phat2}
\hat{p}_j                                  =\frac{(x_{j-1}p_j)p_j-p_j^2
x_{j-1}}{\sqrt{(p_jx_{j-1})^2-p_j^2x_{j-1}^2}}
\end{eqnarray}
The sign  $\epsilon_j$ should be  positive or negative  depending upon
whether  $m^2/z_j$ is  larger or  smaller than  $z_jx_{j-1}^2$  (it is
useful to note that $2(p_jx_{j-1})=(m^2/z_j+z_jx_{j-1}^2)$). Therefore
if we prescribe  the first vertex vector $x_0$  (this is always chosen
in our  process as the  original $q$ energy-momentum vector)  then the
directrix vectors  as well as the vertices  are determined recursively
up to a sign:
\begin{eqnarray}
k_j   =  p_j   +  \epsilon_j\hat{p}_j   \nonumber\\   x_{j}=x_{j-1}  +
\epsilon_j\hat{p}_j
\end{eqnarray}
It is  evident that  the other sign  will determine  the corresponding
$\ell_j$. We note however, that neither the necessary rank ordering of
the  hadrons, nor  the colour  ordering of  the directrix  gluons, are
experimental observables. Therefore in this form the result has solely
a  theoretical meaning.   It is  also necessary  to take  the possible
transverse fluctuations in the fragmentation process (mentioned before
as stemming from tunneling) into account before any observables can be
presented.

\section{Results}
\label{results}
In the  previous section, we have  proposed a method  of fragmenting a
multigluon string based  upon the Area Law. It can  be described as an
iterative stochastic process along  the directrix curve, which fulfils
the  Area  Law at  every  step.  We  have  discussed  a few  different
approaches for  passing a gluon corner,  cf section \ref{gluoncorner},
and from among these we have  picked a preferred one.  In this section
we present  a few  basic results from  this preferred solution  and we
also  compare  it  with  JETSET  and  the  other  approaches  we  have
discussed.

\begin{figure}[!t]
\begin{center}%\vspace{-1cm}
\epsfig{figure=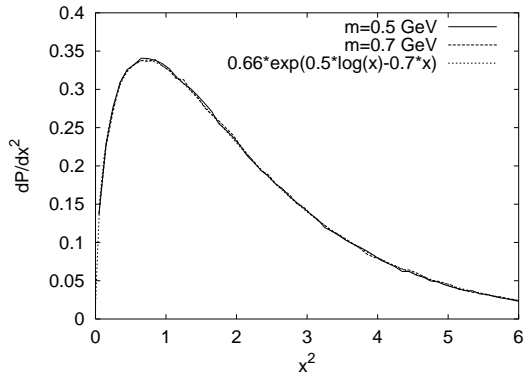, width=0.6\textwidth }
\caption{The  $x^2$ distribution  of  the production  vertices in  our
method for different  values of the hadron mass,  together with a plot
of  the  $\Gamma$  distribution  in Eq.(\ref{gamma})  with  a=0.5  and
b=0.7.} \label{x2dist}
\end{center}
\end{figure}

We start with comparing the  distribution of $x^2$ from our model with
 the distribution  of the proper  time of the break-up  points derived
 inside the $(1+1)$-dimensional model \cite{BA}.
\begin{eqnarray}
\label{gamma}
H(\Gamma)=N \Gamma^{a} exp(-b\Gamma)
\end{eqnarray}
The  result is  presented  in Fig.\ref{x2dist}.  As  mentioned in  the
discussion at the end of  section 5.2, sometimes certain values of $z$
can not  be accomodated  when we pass  the corner according  to choice
C3. Our Monte Carlo simulation  program handles this by starting a new
event from  the beginning whenever  this occurs.  One may  then expect
that this might cause changes  in the $z$-distribution and also in the
$\Gamma$-distribution that  we obtain. This is, however,  not the case
as can be seen from Fig.(14) and Fig.(15). Further these distributions
are  as they  should  be according  to  the Lund  Model formulae.  The
$\Gamma$-distribution is completely independent of the mass value used
just as in the standard, $1+1$-dimensional Lund Model.

\begin{figure}[!h]
\begin{center}%\vspace{-1cm}
\epsfig{figure=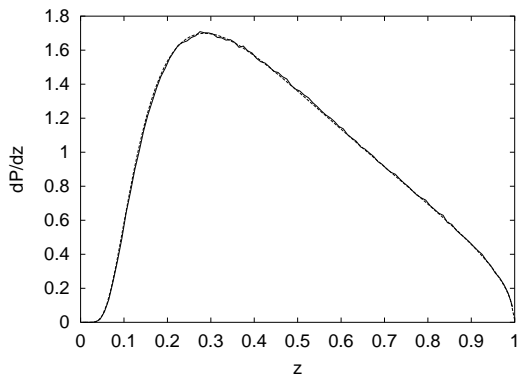, width=0.6\textwidth }
\caption{ z  values used in  particle production in our  method (solid
line)  compared   to  the  Lund   fragmentation  function  f(z)(dotted
line).  Though z  value generation  is according  to  the distribution
f(z), not every value produced  is used. This plot shows however, that
the  rejection  of z  values  in the  present  method  is unbiased.  }
\label{zdist}
\end{center}
\end{figure}

To  examine   further  effects   in  connection  with   our  ``restart
strategy'',  we  define  the  following  measure $\Delta  y$,  of  the
correlation between two adjacent particles in the production process
\begin{eqnarray}
\label{deltayeq}
\Delta y=ln( \frac{z_{1}}{z_{2}(1-z_{1})})
\end{eqnarray}
It corresponds to the difference  in rapidity between the particles in
the $(1+1)$-dimensional  model. This distribution would  be altered if
we  rejected particular  correlations between  successive values  of z
(e.x. a  large value followed  by a small  value etc) more  often than
others.

\pagebreak

\begin{figure}[h]
\begin{center}
\epsfig{figure=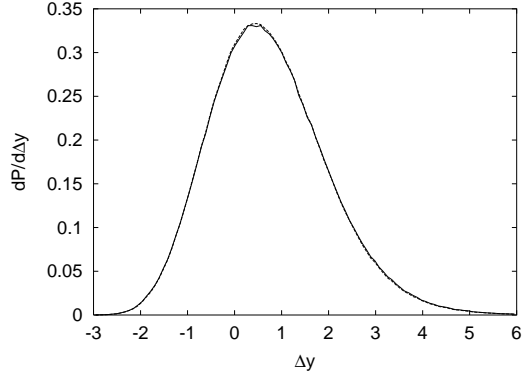, width=0.6\textwidth}
\end{center}
\caption{The $\Delta y$  distribution, defined in eq.(\ref{deltayeq}),
from our  method (solid line) compared with  the distribution obtained
using   only  the  Lund   symmetric  fragmentation   function  (dashed
line).  } \label{deltay}
\end{figure}
   
\begin{figure}[h]
\begin{center}%\vspace{-1cm}
\epsfig{figure=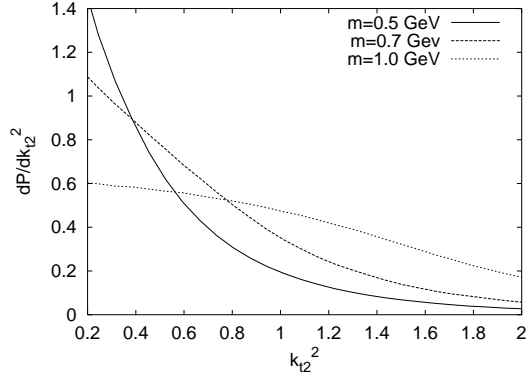,width=0.6\textwidth }
\end{center}
\caption{Plot   of    $\frac{dP}{d   (   k_{    \perp   2}^{2})}$   vs
$k_{\perp2}^{2}$  measured   in  $(GeV)^{2}$.  Modifications   of  the
directrix are on the scale of hadronic mass.} \label{kt2dist}
\end{figure}
\pagebreak
In Fig.\ref{deltay} we compare our  result with the result one obtains
using  only the  Lund symmetric  fragmentation function.  There  is no
noticeable  difference between  the  two curves,  suggesting that  the
restart procedure  (involving possible rejection of some  z values) is
effectively insensitive to such correlations between adjacent values.

In Fig.\ref{kt2dist} we show the distribution of our ''local bending''
parameter,  $k_{\perp2}^{2}$,  defined  in eq.(\ref{kperp2def}).  This
figure  illustrates that the  modifications of  the directrix  that we
make when  we apply our method are  of the order of  the hadronic mass
scale. For values of $k_{\perp2}^{2} > m^2$ there is a fall off faster
than a gaussian.

\begin{figure}[h]
\begin{center}\vspace{1cm}
\mbox{\epsfig{figure=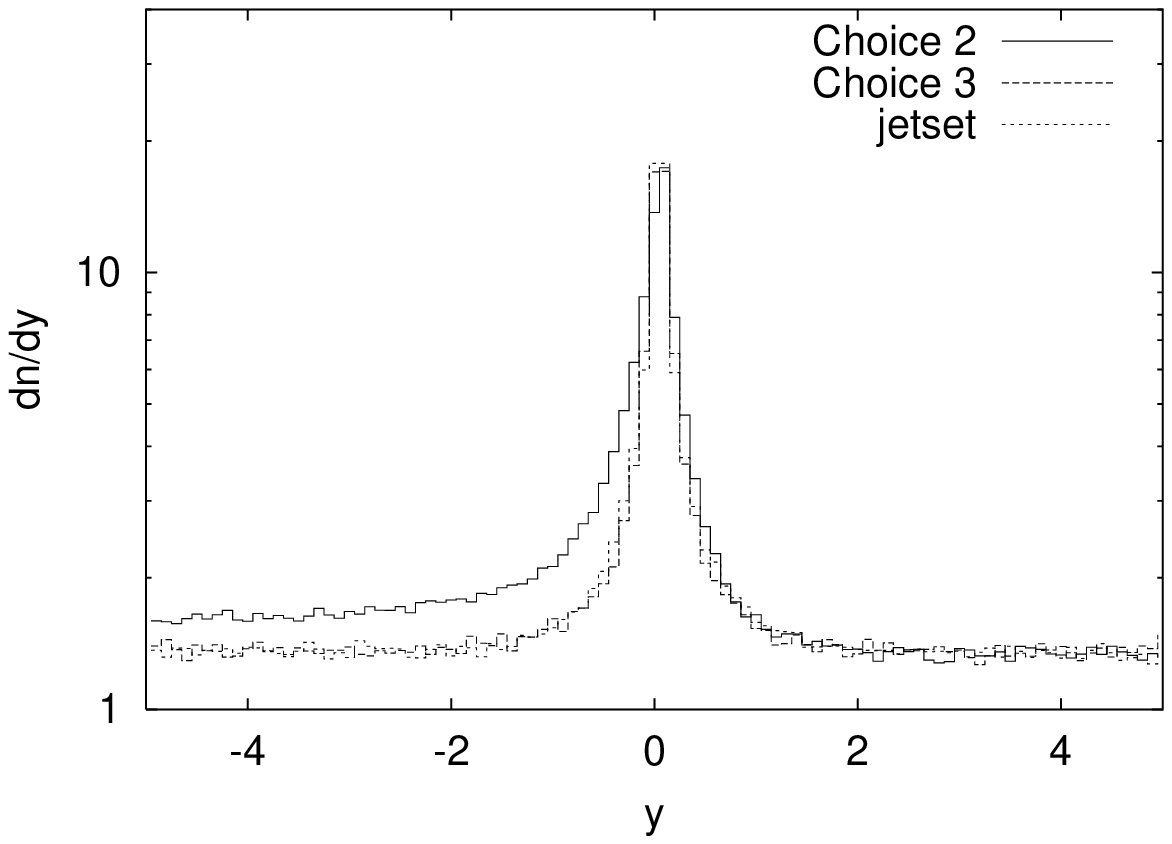,                    width=0.5\textwidth
}\epsfig{figure=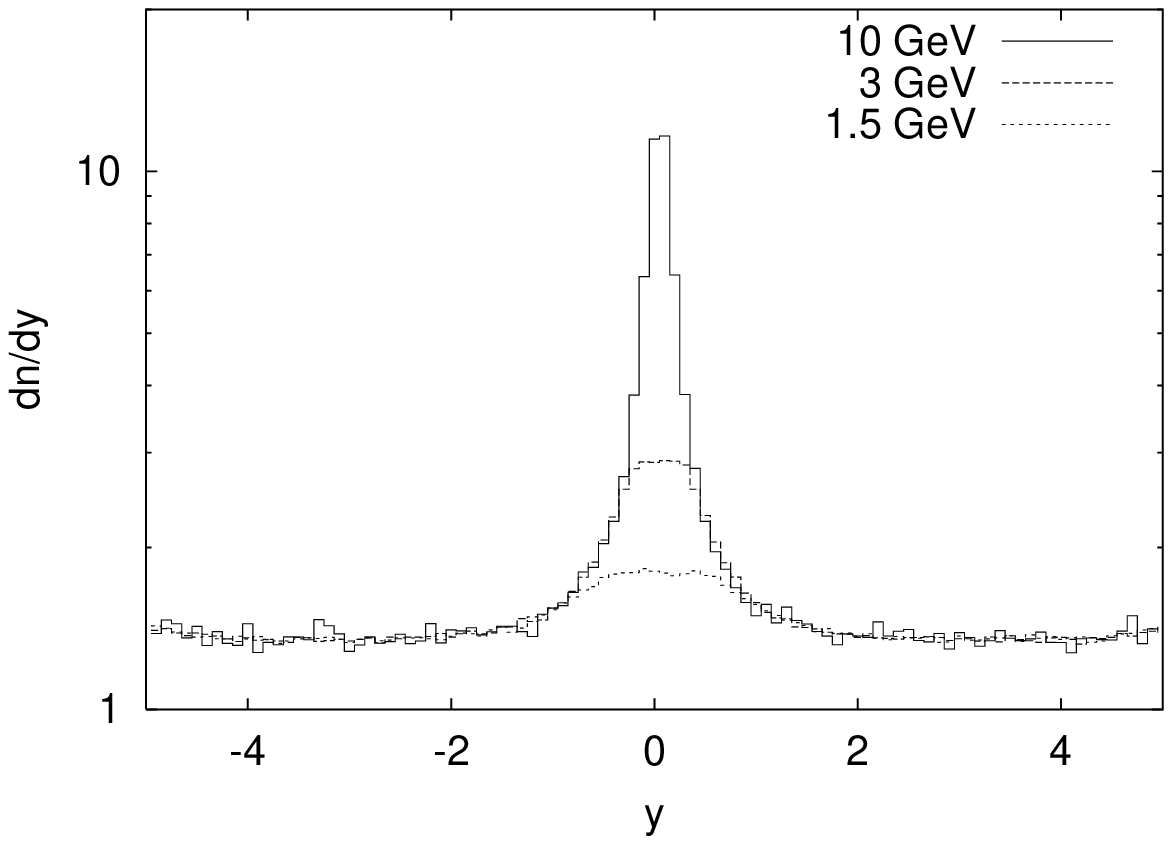,width=0.5\textwidth }}
\end{center}
\caption{The  figure  to the  left  shows  rapidity distributions  for
different  choices for the  vector k.  The figure  to the  right shows
rapidity distributions for a 3  jet event, obtained with the preferred
method (C3),  for different  gluon energies. Note  that the y  axis is
plotted on a log scale here. }
\label{rapidity15}
\end{figure}
In the  left-hand side plot  of Fig.\ref{rapidity15} we  summarize the
rapidity  distributions obtained  using the  different  approaches for
passing   a   gluon  corner,   C2   and   C3   described  in   section
\ref{gluoncorner}, and  compare them to a curve  produced using PYTHIA
in one simple  situation: a perturbative string state  consisting of a
quark, an antiquark and a single gluon. We conclude that modifying the
directrix leads to  the production of too many  particles over a large
range  of rapidity  if  the  modifications are  not  "smooth". In  the
right-hand  side  of  Fig.\ref{rapidity15}  we  show  a  few  rapidity
distributions obtained  from using our preferred  method for different
energies of the gluon.

\pagebreak
\begin{figure}[h]
\begin{center}\vspace{2cm}
\mbox{\epsfig{figure=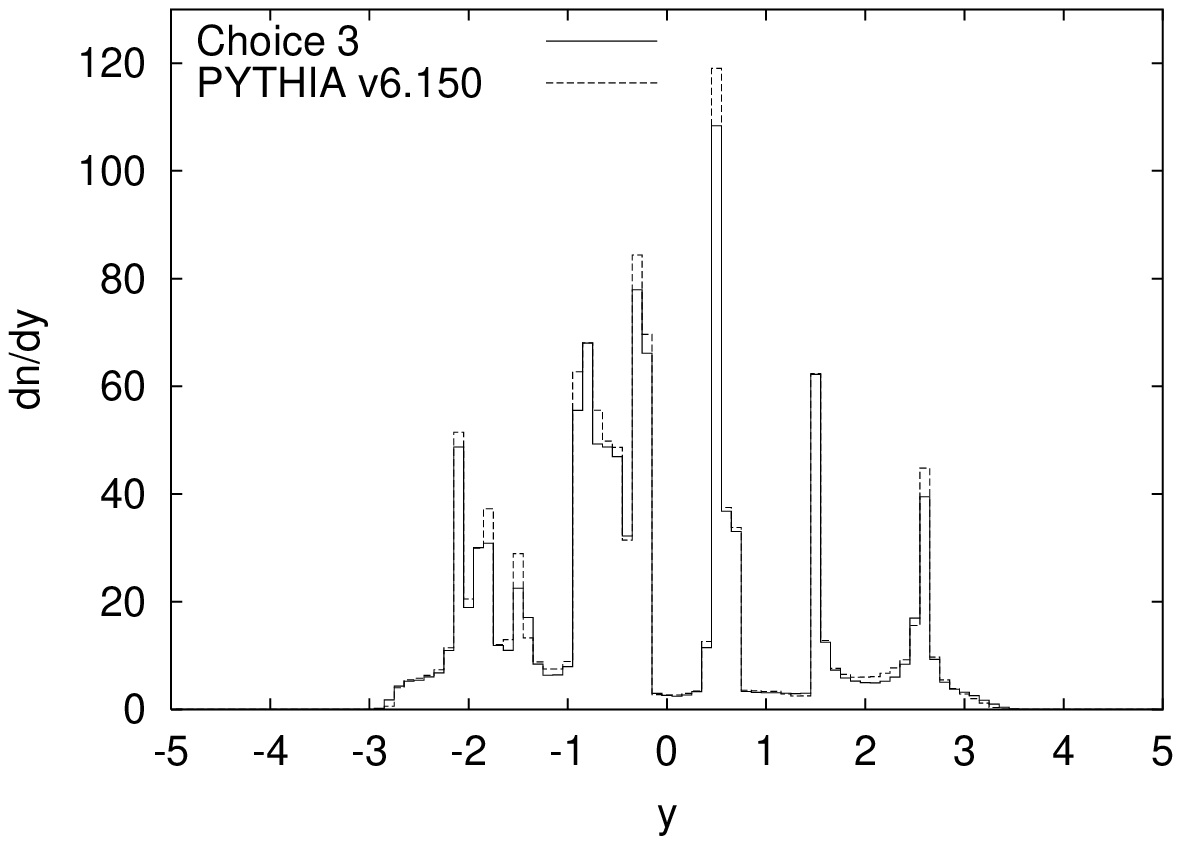,
width=0.5\textwidth}\epsfig{figure=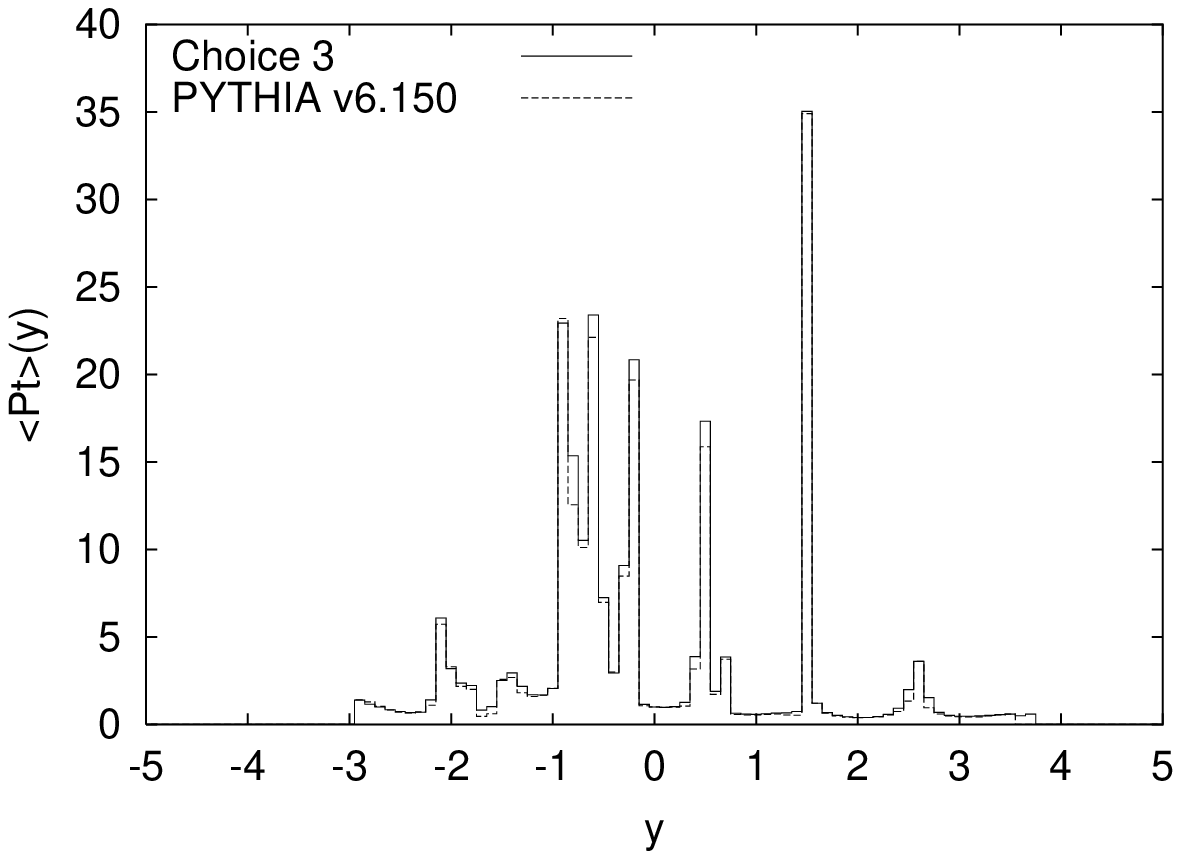,width=0.5\textwidth}}
\caption{    The    rapidity     distribution    (left)    and    mean
$|p_{\perp}^{hadrons}|$ (GeV) as a function of rapidity (right), for a
string state  of $\lambda$  measure 70 (which  is above the  average $
\lambda $ for  the energy and $p_{\perp}^{cut}$ used  for this plot: $
\sqrt{s}=2000$ GeV , $p_{\perp}^{cut} = 2 GeV$ ) generated by ARIADNE,
fragmented using our method (solid curve) and PYTHIA (dashed curve).}
\label{multijet}
\end{center}
\end{figure}

In  Fig.(\ref{multijet})  we show  an  example  of inclusive  rapidity
distributions  obtained  by  JETSET/PYTHIA  and our  method.  For  the
comparison, we  have used an  arbitrarily chosen partonic  event taken
from ARIADNE. A close examination will show that in general there is a
small multiplicity difference  in the gluon jets. We  have traced that
to  the  fact  that  the  two  to three  particles  with  the  largest
energy-momenta in the jets are faster according to our method compared
to JETSET.  It is impossible  to compare to experimental  data because
firstly  in this  case we  have  only made  use of  a single  partonic
event.  And   secondly,  at  present   we  have  not   introduced  the
``fragmentation transverse  momenta''.  Nevertheless we  conclude that
the inclusive distributions are very similar.

\pagebreak

\begin{figure}[h]
\begin{center}\vspace{1cm}
\epsfig{file=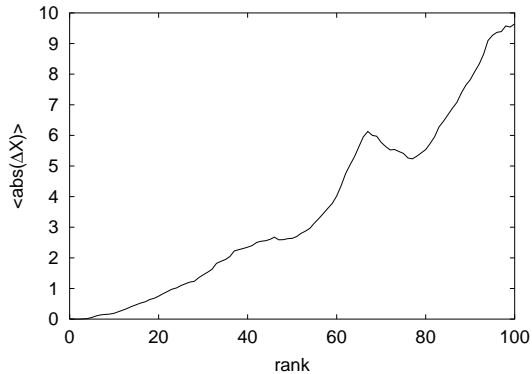, width=0.6\textwidth}
\end{center}
\caption{Comparison  of JETSET  and our  method on  an event  to event
basis:  We  plot  here  the  average  of the  absolute  value  of  the
difference  (GeV) between the  hadronic curves  for the  same partonic
event used in Fig.(\ref{multijet}),  obtained from the two programs as
a function of  rank, where for every event the  particles of a certain
rank  were produced  using exactly  the  same value  of z  in the  two
alternative procedures. } \label{exclusive}
\end{figure}

But we are definitely not dealing  with the same particles on an event
to event  basis. To  show this,  we compare the  two procedures  on an
exclusive, event to event  basis in Fig.\ref{exclusive}. For this plot
we  devised a  procedure  to use  exactly  the same  values  of z  for
particle production  in both programs.  Since both  our procedure, and
PYTHIA might need to reject some  sequences from time to time, we took
only  those events  where no  such  rejection was  required by  either
program. We have plotted the distribution of the absolute value of the
difference  of  the   two  hadronic  curves,  $  \sqrt{|\sum_{j=1}^{n}
(p_{C3})_{j} -\sum_{j=1}^{n} (p_{PYTHIA})_{j}|^{2} } $ , as a function
of rank.   The curve clearly  shows that the hadronic  curves obtained
from  the two  programs  are different.  The  absolute value  prevents
cancellation of the differences which would occur in an average over a
large  number  of  events, as  we  find  from  the comparison  of  the
inclusive distributions.

\section{Concluding Remarks}
\label{remarks}
We  have presented a  precise method  to implement  the Lund  Area Law
(Eq. (\ref{area}))  to fragment a  multigluon string state.  The final
state  hadrons are produced  in an  iterative stochastic  process. The
energy-momentum vectors $\{p_j\}$  can be laid out in  rank order as a
curve, the $X$-curve,  with a vertex vector at  every point connecting
to the directrix curve. The directrix corresponds to the parton energy
momentum vectors  laid out  in colour order.  It describes on  the one
hand the orbit of a  (massless) $q$-particle connected to a string and
on the other hand the whole string surface.

It is possible  to describe the whole process  by analytical means, in
particular in terms of a  transfer matrix formalism similar to the one
used for  the $(1+1)$-dimensional model in \cite{BAFS}.   This time it
is necessary to  make use of a $(1+3)$-dimensional  framework. We will
investigate this problem in the future.

In  this  paper, we  have  neglected  the  possibility of  introducing
transverse   fluctuations   in   the   hadronisation   process.   Such
fluctuations are  introduced on the  basis of tunnelling  arguments in
the $(1+1)$-dimensional model.  We feel  that they are necessary for a
consistent quantum  mechanical treatment (there is no  way to localise
the string  surface area better  than what is allowed  by Heisenberg's
indeterminacy  relations).  We  will   examine  the  effects  of  such
transverse fluctuations on our results in the future.

The  directrix curve  is in  general given  by a  perturbative cascade
(although   in   some  cases   precise   matrix   elements  are   also
available). This  means that the  partonic states we  are fragmenting,
are resolved  only down  to some cutoff,  usually in virtuality  or in
terms of the smallest allowed partonic excitations. String dynamics is
infrared stable in the sense that minor modifications of the directrix
only have a  local influence, due to the minimal  nature of the string
world surface. We  have made use of this  freedom in the fragmentation
process and  in this way the  partonic states will be  defined down to
the  hadronic mass-scale. We  will investigate  the properties  of the
soft gluons thus introduced into the partonic state, in the future.

We have shown that the  fragmentation process, with step size equal to
the hadronic mass can be defined also in the limit of a vanishing mass
in terms  of a differential  process. The corresponding  solution, the
${\cal  P}$-curve, is stretched  in between  two curves,  the original
directrix, ${\cal  A}$, and another curve, the  ${\cal L}$-curve, also
with a  light-like tangent. The  correspondence to the  vertex vectors
for the  fragmentation process are connectors to  the ${\cal P}$-curve
reaching out to the ${\cal A}$  and the ${\cal L}$-curves. The area in
between the ${\cal P}$-curve  and the directrix ${\cal A}$ corresponds
to  a generalised rapidity  variable in  the same  way as  the average
hyperbola defines rapidity for the $(1+1)$-dimensional model. There is
an  interesting   relationship  to  the   group  of  rotations   in  a
$(1+4)$-dimensional  space which  we will  investigate further  in the
future.

We have also shown that there is a duality (with properties similar to
the  Parton-Hadron  Duality introduced  by  the  St. Petersburg  group
\cite{STPETERSBURG})  between  hadronic   $X$-curve,  defined  by  our
process  and  the original  directrix.   We  note,  however, that  the
hadrons produced in our process always stem from the energy-momenta of
two or more  partons. The vertex vectors $x_j$  contain the ''memory''
of the earlier partons. It is nevertheless possible to reconstruct the
directrix  from the  $X$-curve, although  the relationship  contains a
large number of  degrees of freedom (the number  of degrees of freedom
increases further if we  introduce transverse momentum fluctuations in
the fragmentation  process). We  will investigate these  properties in
the future.

The process  we have  defined has also  been implemented into  a Monte
Carlo  simulation program.  Due to  the way  it is  constructed  it is
fairly direct  to include also the  multi-particle production features
of the  original JETSET, i.e.  to include probabilities  for different
$(q\bar{q})$-flavors,  different   kinds  of  mesons,   the  decay  of
resonances etc.  This is necessary for a  comparison with experimental
data.

\section{Acknowledgements}
We  would like  to thank  G. Gustafson  for extensive  discussions and
T.  Sj\"ostrand and  L. L\"onnblad  for a  lot of  help both  with the
material  and in  particular with  the  way the  Monte Carlo  programs
JETSET and ARIADNE work.

%%%%%%%%%%%%%%%%%%%%%%  thebibliography  %%%%%%%%%%%%%%%%%%%%%%%%%%%%

\end{document}